\documentclass[12pt,reqno]{amsart}
\usepackage{fullpage}
\usepackage{amsmath,amssymb}
\usepackage{graphics,graphicx}
\usepackage{amsthm}
\usepackage{mathrsfs}
\usepackage{slashbox}
\usepackage[table]{xcolor}
\usepackage{epstopdf}

\newcommand{\mi}{\ensuremath{\mathrm{i}}}
\newcommand{\me}{\ensuremath{\mathrm{e}}}
\newcommand{\dif}{\ensuremath{\mathrm{d}}}
\newcommand{\pd}{\ensuremath{\partial}}
\newcommand{\Id}{\ensuremath{\mathsf{I}}}
\newcommand{\CC}{\ensuremath{\mathbb{C}}}
\newcommand{\RR}{\ensuremath{\mathbb{R}}}
\newcommand{\NN}{\ensuremath{\mathbb{N}}}

\newcommand{\lam}{\ensuremath{\lambda}}
\newcommand{\eps}{\ensuremath{\epsilon}}

\renewcommand{\vec}[1]{\ensuremath{\mathbf{#1}}}
\newcommand{\mat}[1]{\ensuremath{\mathsf{#1}}}
\newcommand{\beq}{\begin{equation}}
\newcommand{\eeq}{\end{equation}}
\DeclareMathOperator{\sech}{sech}
\DeclareMathOperator{\res}{res}

\newcommand{\spm}{{\ensuremath{{\scriptscriptstyle\pm}}}}
\renewcommand{\sp}{{\ensuremath{{\scriptscriptstyle +}}}}
\newcommand{\sm}{{\ensuremath{{\scriptscriptstyle -}}}}

\numberwithin{equation}{section}

\newtheorem{theorem}{Theorem}
\newtheorem*{conjecture}{Conjecture}

\theoremstyle{definition}
\newtheorem{rhp}{Riemann--Hilbert Problem}

\theoremstyle{remark}

\begin{document}
\title{The Gaussian Semiclassical Soliton Ensemble and Numerical Methods for the Focusing Nonlinear Schr\"odinger Equation}
\author{Long Lee}
\email{llee@uwyo.edu}
\author{Gregory Lyng}
\email{glyng@uwyo.edu}
\author{Irena Vankova}
\email{ivankova@uwyo.edu}
\address{Department of Mathematics, University of Wyoming, Laramie, WY 82071-3036}
\date{updated \today}

\begin{abstract}
We report on a number of careful numerical experiments motivated by the semiclassical (zero-dispersion, $\eps\downarrow 0$) limit of the focusing nonlinear Schr\"odinger equation. Our experiments are designed to study the evolution of a particular family of perturbations of the initial data.
These asymptotically small perturbations are precisely those that result from modifying the initial-data by using formal approximations to the spectrum of the associated spectral problem; such modified data has always been a standard part of the analysis of zero-dispersion limits of integrable systems. However, in the context of the focusing nonlinear Schr\"odinger equation, the ellipticity of the Whitham equations casts some doubt on the validity of this procedure.  To carry out our experiments, we introduce an implicit finite difference scheme for the partial differential equation, and we validate both the proposed scheme and the standard split-step scheme against a numerical implementation of the inverse scattering transform for a special case in which the scattering data is known exactly. As part of this validation, we also investigate the use of the Krasny filter which is sometimes suggested as appropriate for nearly ill-posed problems such as we consider here. Our experiments show that that the $O(\eps)$ rate of convergence of the modified data to the true data is propagated to positive times including times after wave breaking.
\end{abstract}
\maketitle

\section{Introduction}\label{sec:intro}
\subsection{Background}\label{ssec:back}
In their pioneering work, Lax \& Levermore \cite{LL1,LL2,LL3} used the inverse scattering transform (IST) to study the zero-dispersion limit of the initial-value problem for the Korteweg--de Vries equation:
\begin{subequations}\label{eq:kdv_ivp}
\begin{align}
\pd_ty-6y\pd_xy& =\eps^2\pd_x^3y\,, \label{eq:kdv}\\ 
y(x,0) &=y_0(x)\,. \label{eq:kdv_data}
\end{align}
\end{subequations}
That is, they were able to characterize the limiting behavior of the family, indexed by $\eps>0$, of solutions $y^{(\eps)}(x,t)$ of \eqref{eq:kdv_ivp} in the limit $\eps\downarrow 0$. Their results (and those of others who have since extended and refined the analysis of \eqref{eq:kdv_ivp}, e.g., \cite{V85,V87,T,DVZ,CG}) show---as one might guess---that for small times, $y^{(\eps)}(x,t)$ converges strongly to $\bar y(x,t)$, the solution of 
\begin{subequations}\label{eq:hopf_ivp}
\begin{align}
\pd_t\bar y-6\bar y\pd_x\bar y& =0\,, \label{eq:hopf}\\ 
\bar y(x,0) &=y_0(x)\,. \label{eq:hopf_data}
\end{align}
\end{subequations}
For any fixed $\eps>0$, solutions of \eqref{eq:kdv_ivp} with smooth, decaying data exist and remain smooth for all $t>0$. By contrast, the limiting equation \eqref{eq:hopf} is known to support solutions which develop shocks in finite time regardless of the smoothness of the data. Indeed, a major impetus for the study of this problem has been to understand how the dispersive term in \eqref{eq:kdv} prevents shock formation. Roughly, the solution develops rapid nonlinear oscillations which carry energy away from a developing shock, and Lax--Levermore theory provides a rather precise description of the character of these oscillations in various regions of the $xt$-plane. An integral component of this description are the Whitham or modulation equations; these partial differential equations describe the local evolution of the large-scale structures in the solution. Notably, in the case of \eqref{eq:kdv}, these equations are of hyperbolic type.  

The solution of \eqref{eq:kdv_ivp} by IST is intimately connected with the spectrum of the Schr\"odinger operator,
\[
-\eps^2\frac{\dif^2}{\dif x^2}+y_0\,,
\]
and the first step in Lax \& Levermore's analysis was to replace the true spectrum with WKB approximations. This replacement creates a sequence of reflectionless potentials $y_0^{(\eps)}(x)$ which converge to $y_0(x)$ in $L^2(\RR)$ as $\eps\downarrow 0$. Using the IST, Lax \& Levermore were then able to write down essentially explicit representations of the family of solutions of \eqref{eq:kdv} corresponding to this sequence of modified data, and they were able to analyze and describe the limiting structure of this family. 

These ideas and their extensions have also been used to address other problems including the semiclassical limit of the \emph{defocusing} nonlinear Schr\"odinger equation \cite{JLM}, 
the continuum limit of the Toda lattice \cite{DM}, and a continuum limit of a discrete nonlinear Schr\"odinger equation \cite{S}. In all of these analyses, a step corresponding to the  modification of the initial data, as described above for \eqref{eq:kdv_ivp}, has been the starting point of the analysis. In each of these cases, the Whitham equations are hyperbolic, hence locally well posed. Thus, in view of the $L^2$-convergence of the modified data to the true data, it seems reasonable to expect that the convergence holds for $t>0$ as well. Here, however, we address a case in which the Whitham equations are \emph{elliptic}, and such an expectation seems much more dubious. Our aim here is to better understand the effect of modifying the data in such a case.

\subsection{Focusing Nonlinear Schr\"odinger Equation and Semiclassical Limit}\label{ssec:fnls}
We consider the initial-value problem for the semiclassically-scaled focusing nonlinear Schr\"odinger (NLS) equation:
\begin{subequations}\label{eq:nls_ivp}
\begin{align}
\mi\eps \pd_t u+\frac{\epsilon^2}{2}\pd_x^2u&+|u|^2u=0\,,\label{eq:nls}\\
u(x,0)&=u_0(x)\,.\label{eq:data}
\end{align}
\end{subequations}
Equation \eqref{eq:nls} is a universal model equation that arises in models of diverse physical scenarios; it describes the envelope dynamics of a monochromatic wave in a weakly dispersive nonlinear medium in which diffusive effects are negligible  \cite{APT,M_book,SS}. For example, it is a simple model for the propagation of light in optical fibers \cite{A}. In \eqref{eq:nls}, $0<\epsilon\ll1$ is a constant parameter which measures the ratio of dispersion to nonlinearity. 

Our interest is the \emph{semiclassical} or \emph{zero-dispersion} limit of \eqref{eq:nls_ivp}. That is, we suppose that the initial data $u_0$ is fixed, and we solve \eqref{eq:nls_ivp} for each small $\epsilon>0$. We describe below our assumptions on $u_0$ which guarantee the existence of a unique global solution to \eqref{eq:nls_ivp} so that, in principle at least, this first step is possible. Then, given the resulting family (indexed by $\epsilon$) of solutions,
\[
u(x,t)=u^{(\epsilon)}(x,t)\,,
\]
the goal is to describe the asymptotic behavior of these solutions in the limit $\epsilon\downarrow 0$.
The first breakthrough for this problem was due to Kamvissis, McLaughlin, \& Miller~\cite{KMM} for initial data of the form
\begin{equation}
u_0(x)=A_0(x)\,,
\label{eq:amplitude}
\end{equation}
where $A_0:\mathbb{R}\to(0,A]$ is even, bell-shaped, and real analytic. More precisely, $A_0$ is assumed to 
\begin{enumerate}
\item[(i)] decay rapidly at $\pm\infty$;
\item[(ii)] be an even function, i.e., $A_0(x)=A_0(-x)$ for all $x\in\mathbb{R}$;
\item[(iii)] have a single genuine maximum at $x=0$, i.e., $A_0(0)=A$, $A_0'(0)=0$, $A_0''(0)<0$; and 
\item[(iv)] be real-analytic.
\end{enumerate}
Henceforth, we adopt these assumptions. 

We remark that for fixed $\epsilon>0$, well-posedness for the Cauchy problem (with $\epsilon$-independent data, as in \eqref{eq:data}) is well known. For example, we note that Ginibre \& Velo \cite{GV} have shown that if 
\(
u_0\in H^1(\RR)\cap L^\infty(\RR),
\)
then \eqref{eq:nls} has a unique global solution $u(t)$ in $\mathscr{C}(\RR; H^1(\RR)\cap L^\infty(\RR))$; the solution depends continuously on the data.
Moreover, in the case (as we consider here) that $u_0\in\mathscr{S}(\RR)$---the Schwartz space of rapidly decaying functions, it is known that $u(\cdot,t)$ is also in $\mathscr{S}(\RR)$ for each $t$ \cite{FT}. The issue is that this well-posedness is not uniform in $\eps$ \cite{DGK}. 

\subsection{Inverse scattering transform}
As in the other problems to which Lax--Levermore theory has been applied, equation \eqref{eq:nls} is integrable, i.e., there is an associated Lax pair. Thus, we can solve the initial-value problem \eqref{eq:nls_ivp} by IST; \ref{sec:ist} contains an outline of the process. Indeed, it could be argued that the integrability of \eqref{eq:nls} is the only feature that makes the task of obtaining (postbreak) asymptotics even appear tractable. The equivalent problem for nonintegrable variants of \eqref{eq:nls} appears to be widely open \cite{C}.

The first step, then, in solving \eqref{eq:nls_ivp} is an analysis of the nonselfadjoint Zakharov--Shabat eigenvalue problem (one half of the Lax pair for \eqref{eq:nls}):
\begin{equation}\label{eq:zs}
\epsilon\frac{\dif}{\dif x}
\begin{bmatrix}
w_1(x;\lambda) \\
w_2(x;\lambda)
\end{bmatrix} =
\begin{bmatrix}
-\mi\lambda & A_0(x) \\
-A_0(x) & \mi\lambda
\end{bmatrix}\begin{bmatrix}
w_1(x;\lambda) \\
w_2(x;\lambda)
\end{bmatrix}.
\end{equation}
In \eqref{eq:zs}, $w_1$ and $w_2$ are auxiliary functions and $\lambda\in\mathbb{C}$ is a spectral parameter.  For each $\epsilon>0$ and for $A_0$ as described above, it is known (see \cite{KS}) that the discrete spectrum of \eqref{eq:zs} is confined to the imaginary axis. Beyond this, a formal WKB method applied to \eqref{eq:zs} suggests that the reflection coefficient is small beyond all orders and the imaginary eigenvalues are given by a quantization condition of Bohr--Sommerfeld type. Since precise information about the true scattering data (discrete eigenvalues of \eqref{eq:zs} and the reflection coefficient) is not known, a natural way forward---following Lax \& Levermore---is to use the (formal) WKB scattering data in its place. For each small $\epsilon>0$ this procedure, neglecting the reflection coefficient and using the WKB eigenvalues, amounts to replacing the true initial data $A_0$ with some other initial condition $u_0^{(\epsilon)}$ which depends on $\epsilon$ and for which the WKB spectral data is the \emph{true} spectral data. Because we neglect reflection, each solution of \eqref{eq:nls} with initial data $u_0^{(\epsilon)}$ is an $N$-soliton with $N\sim\epsilon^{-1}$. The collection of all these exact $N$-soliton solutions of \eqref{eq:nls} (with $N\to\infty$ and $\epsilon\downarrow 0$) is called the \emph{semiclassical soliton ensemble} (SSE) associated with $A_0$.  

The analysis of Kamvissis et al. \cite{KMM} is almost wholly focused on the inverse scattering step for SSEs. Now, there are at least two special cases for the which the spectral data is known exactly. For $A_0(x)=A\sech(x)$, Satsuma \& Yajima \cite{SY}, after a clever transformation, have shown how to write down explicit formulae for the eigenvalues, proportionality constants, and reflection coefficient coming from \eqref{eq:zs}. More recently, Tovbis \& Venakides \cite{TV} introduced a special family of initial data (with a complex phase) for which the forward scattering problem can also be treated exactly. This family of data forms the foundation of the related work on the semiclassical limit by Tovbis, Venakides, \& Zhou \cite{TVZ}. It is clearly of interest to obtain results for a much wider class of initial data, and a natural first step is to look at the bell-shaped data that generate SSEs. 

Strictly speaking, the analysis of Kamvissis et al. \cite{KMM} describes the asymptotic behavior of such SSEs for $t\neq 0$. At $t=0$ there is the complementary result of Miller:
\begin{theorem}[Miller\,\cite{M}]\label{thm:miller}
In the situation described above, there is a sequence $(\epsilon_N)_{N=1}^\infty$ such that 
\begin{equation}
\lim_{N\to\infty}\epsilon_N=0,
\end{equation}
and such that for each $x\neq 0$ there exists a $K_x$ such that 
\begin{equation}
|u_0^{(\epsilon_N)}(x)-A_0(x)|\leq K_x\epsilon_N^{1/7-\nu},\quad N=1,2,3,\ldots
\end{equation}
for all $\nu>0$.
\end{theorem}

As noted above, a fundamental issue that remains unresolved is to connect the asymptotics of the SSE to those of the true solution of the initial-value problem \eqref{eq:nls_ivp}. That is, Theorem~\ref{thm:miller} shows that that SSE and $A_0$ are asymptotically pointwise close at $t=0$. However, \eqref{eq:nls} has modulational instabilities whose exponential growth rates become arbitrarily large in the semiclassical limit---the Whitham equations are elliptic. Thus, it is not possible to conclude from Theorem \ref{thm:miller} that any member of the SSE and the corresponding true solution are close for any $t>0$. 
To attack this difficulty, one could try to rigorously estimate the deviation of the WKB spectral data corresponding to $A_0$ from $A_0$'s true spectral data. With such eigenvalue-by-eigenvalue control in hand, one could then try to incorporate this information into the asymptotic analysis. This is the ongoing work of \cite{BMM}. Our complementary goal in this paper is to better understand, by numerical experiment, the relationship between the SSE and $A_0$ at $t=0$ and between the SSE and the true solution $u(x,t)$ for $t>0$. In particular, our experiments support the following conjecture.

\begin{conjecture}
For small times, despite the presence of modulational instability, the particular asymptotically small modification of the initial data used by Kamvissis et al. to generate a semiclassical soliton ensemble for bell-shaped data does not affect the limiting behavior. 
\end{conjecture}
In the remainder of this paper, we describe the process we used to generate numerical evidence that supports this conjecture. Remarkably, our experiments show that the $O(\epsilon)$ rate of convergence of the modified data to the true data is propagated to positive times, including times after wave breaking.

\subsection{Plan}
To aid the reader, we now outline the contents of the remainder of this paper. In Section \ref{sec:istwkb} we describe the machinery from the theory of integrable systems necessary to obtain an $N$-soliton solution of \eqref{eq:nls} with initial data of the form \eqref{eq:amplitude}. That is, assuming that  reflectionless scattering data for Zakharov--Shabat problem \eqref{eq:zs} is known, we recall that the solution of equation \eqref{eq:nls} can be obtained by solving an appropriate Riemann--Hilbert problem. In this section, for initial data $A_0(x)=\exp(-x^2)$, we also describe the computation of the WKB eigenvalues for \eqref{eq:zs}. With high-precision approximations of these eigenvalues in hand, we are able to use known techniques \cite{MK,LM} to generate members of the corresponding SSE. 
Sections \ref{sec:methods} and \ref{sec:experiments} are devoted the development and testing of the numerical methods we use for our eventual comparison between members of the Gaussian SSE and the (numerically computed) true evolution of \eqref{eq:nls_ivp}. We use the members of the Satsuma--Yajima ensemble, known exact $N$-soliton solutions, to validate our numerical methods in the range of $\eps$ and $t$ that we consider here. 
Using the methods of Section \ref{sec:methods}, we report on the principal experiment of the paper in  Section \ref{sec:gauss}; this is the aforementioned comparison of true evolution for various values of $\eps$ with that of the corresponding members of the Gaussian SSE. Finally, Section \ref{sec:discuss} contains a discussion of our results. For example, we contrast our results with some examples in the literature which suggest that when $\eps$ is small, equation \eqref{eq:nls} is extremely sensitive to (rough) perturbations of the initial data. \ref{sec:ist} contains an outline of the features of the inverse scattering transform for \eqref{eq:nls} that are used in this paper. 

\section{Riemann--Hilbert and WKB}\label{sec:istwkb}

\subsection{Riemann--Hilbert formulation}
We take as our starting point the fact that every $N$-soliton solution of the focusing NLS equation can be characterized as the solution of a meromorphic Riemann--Hilbert Problem (RHP) with no jumps. The solution of the RHP is a matrix-valued rational function of $\lambda\in\CC$; the solution depends on a set of discrete data---a collection of $N$ complex numbers in the upper-half plane 
\beq\label{eq:evals}
\{\lambda_{N,0},\lambda_{N,1},\ldots,\lambda_{N,N-1}\}\,,
\eeq
$N$ nonzero constants 
\beq\label{eq:propconst}
\{\gamma_{N,0},\gamma_{N,1},\ldots,\gamma_{N,N-1}\}\,,
\eeq
and a choice of $J=\pm1$. One seeks to solve the following problem.
\begin{rhp}\label{rhp:1}
Find a $2\times 2$ matrix-valued function $\mat{m}(\lambda;x,t)$ with the following properties.
\begin{enumerate}
\item $\mat{m}(\lambda;x,t)\to\Id$ as $\lambda\to\infty$.
\item $\mat{m}(\lambda;x,t)$ is a rational function of $\lambda$ with poles confined to the values $\lambda_{N,k}$ and $\lambda_{N,k}^*$. At the singularities, 
\begin{align}
\res_{\lambda=\lambda_{N,k}}\mat{m}(\lambda)& =\lim_{\lambda\to\lambda_{N,k}}\mat{m}(\lambda)\sigma_1^{\frac{1-J}{2}}\begin{bmatrix}   0 & 0 \\ c_{N,k}(x,t) & 0\end{bmatrix}\sigma_1^{\frac{1-J}{2}}\,, \\
\res_{\lambda=\lambda_{N,k}^*}\mat{m}(\lambda)& =\lim_{\lambda\to\lambda_{N,k}^*}\mat{m}(\lambda)\sigma_1^{\frac{1-J}{2}}\begin{bmatrix}   0 & -c_{N,k}(x,t)^* \\ 0 & 0\end{bmatrix}\sigma_1^{\frac{1-J}{2}}\,.
\end{align}
Here, 
\beq
c_{N,k}(x,t):=\left(\frac{1}{\gamma_k}\right)^J\,\frac{\displaystyle{\prod_{n=0}^{N-1}}(\lambda_{N,k}-\lambda_{N,n}^*)}{\displaystyle{\prod_{\genfrac{}{}{0pt}{}{n=0}{n\neq k}}^{N-1}}(\lambda_{N,k}-\lambda_{N,n})}\exp\left(\frac{2\mi J(\lambda_{N,k}x+\lambda_{N,k}t^2)}{\epsilon}\right)\,,
\eeq
and $\sigma_1$ is the Pauli matrix 
\[
\sigma_1=\begin{bmatrix} 0 & 1 \\ 1 & 0 \end{bmatrix}\,.
\]
\end{enumerate}
Finally, once the solution of RHP \ref{rhp:1} is found, one recovers an $N$-soliton solution via the formula
\[
u(x,t)=2\mi\lim_{\lambda\to\infty}\lambda m_{12}(\lambda;x,t)\,.
\]
\end{rhp}

As pointed out by Kamvissis et al. \cite{KMM}, the RHP above can be recognized as a classical Pad\'e multipoint interpolation problem. They used RHP \ref{rhp:1} as the starting point of their analysis; as a first step they exchanged the meromorphic problem above for a sectionally holomorphic one. Then, as the result of a substantial amount of work, they were able to transform the sectionally holomorphic RHP to one which is amenable to the steepest-descent techniques of Deift \& Zhou \cite{DZ}. The results of this elaborate analysis are detailed asymptotic formulae for the small-$\eps$ behavior.

We proceed in a different fashion. After making a partial fractions ansatz, it is possible to reduce the solution of RHP~\ref{rhp:1} to the solution of an $N\times N$ linear system; see \cite{MK,LM} or \ref{sec:ist}. Then, given eigenvalues $\{\lambda_{N,j}\}$, constants $\{\gamma_{N,j}\}$, and a pair $(x,t)$ (these appear in the linear system as parameters), we may recover the $N$-soliton solution of \eqref{eq:nls} at $(x,t)$. Thus, to construct members of a SSE associated to initial data $A_0$, we need to compute the WKB eigenvalues of the Zakharov--Shabat problem with $A_0$ appearing as potential. With these in hand, we may then turn to solving the poorly conditioned linear system that is born of RHP~\ref{rhp:1}. We describe the calculation of the WKB eigenvalues in the next section.

\subsection{The WKB Formulae}\label{sec:wkb}
\subsubsection{General Case: Bell-shaped Data}\label{ssec:wkb_general}
\begin{figure}[ht] 
   \centering
   \includegraphics[width=3in]{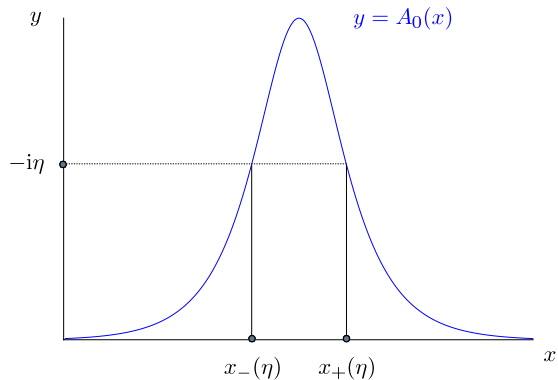} 
   \caption{The turning points $x_\pm(\eta)$.}
   \label{fig:tp}
\end{figure}

We begin by recalling the formulae for the WKB eigenvalues of \eqref{eq:zs}; for more details see \cite{EJLM}. The basic object of interest is the density function
\begin{equation}
\rho^0(\eta):=\frac{\eta}{\pi}\int_{x_-(\eta)}^{x_+(\eta)}\frac{\dif x}{\sqrt{A_0(x)^2+\eta^2}}
=\frac{1}{\pi}\frac{\dif}{\dif\eta}\int_{x_-(\eta)}^{x_+(\eta)}\sqrt{A_0(x)^2+\eta^2}\,\dif x\,,
\label{eq:rho0}
\end{equation}
defined for $\eta\in(0,\mi A)$ where $x_\pm(\eta)$ are the two real turning points; see Figure \ref{fig:tp}.
From $\rho^0$ we obtain the function 
\beq\label{eq:theta0}
\theta^0(\lambda):=-\pi\int_\lambda^{\mi A}\rho^0(\eta)\,\dif\eta
\eeq
which measures the number of WKB eigenvalues on the imaginary axis between $\lam$ and $\mi A$.
We then define for $N=1,2,3,\ldots$
\begin{align}
\epsilon_N&:=-\frac{1}{N}\int_0^{\mi A}\rho^0(\eta)\,\dif\eta \label{eq:hbarn} 
=\frac{1}{\pi N}\int_{-\infty}^\infty A_0(x)\,\dif x\,. \nonumber
\end{align}
Finally, the WKB eigenvalues $\tilde\lambda_{N,k}$ are defined (there are $N$ of them for  $\epsilon_N$) by the formula
\begin{align}
-\int_{\tilde\lambda_{N,k}}^{\mi A}\rho^0(\eta)\,\dif \eta &=\epsilon_N\left(k+\frac{1}{2}\right)\\
&=\frac{\theta^0(\tilde\lambda_{N,k})}{\pi}\,,\quad k=0,\ldots, N-1\,.
\end{align} 
Using the above formulae we may ``simplify'' the left-hand side:
\begin{align*}
-\int_{\tilde\lambda_{N,k}}^{\mi A}\rho^0(\eta)\,\dif \eta&\stackrel{\eqref{eq:rho0}}{=}
-\int_{\tilde\lambda_{N,k}}^{\mi A}\frac{1}{\pi}\frac{\dif}{\dif\eta}\left[\int_{x_-(\eta)}^{x_+(\eta)}\sqrt{A_0(x)^2+\eta^2}\,\dif x\right]\,\dif\eta  \\
&=\frac{2}{\pi}\int_{0}^{x_+(\tilde\lambda_{N,k})}\sqrt{A_0(x)^2+\tilde\lambda_{N,k}^2}\,\dif x\,.
\end{align*}
Therefore, writing $\tilde\lambda_{N,k}=\mi t_{N,k}$ for $t_{N,k}\in(0,A)\subset\mathbb{R}$, we desire to solve the equation 
\begin{equation}
\int_0^{x_+(\mi t_{N,k})}\sqrt{A_0(x)^2-t_{N,k}^2}\,\dif x=\frac{\pi\epsilon_N}{2}\left(k+\frac{1}{2}\right),\;\;k=0,1,2,\ldots N-1\,.
\label{eq:wkbeval}
\end{equation}
In this case, the auxiliary scattering data (proportionality constants) are given by 
\beq\label{eq:gammak}
\tilde\gamma_{N,k}=(-1)^{k+1}\,.
\eeq

\subsubsection{The Gaussian SSE}\label{ssec:gaussian_SSE}
For our numerical experiments, we restrict ourselves to the Gaussian SSE. That is, from now on, we consider the problem \eqref{eq:nls_ivp} with fixed initial data given by 
\begin{equation}
u_0(x)=A_0(x)=\me^{-x^2}.
\label{eq:gauss}
\end{equation}
Then, from \eqref{eq:hbarn}
\begin{equation}\label{eq:epsilon_N}
\epsilon_N=\frac{1}{\pi N}\int_{-\infty}^\infty\me^{-x^2}\,\dif x =\frac{1}{\sqrt{\pi}N}.\,,
\end{equation}
and formula \eqref{eq:wkbeval} becomes
\begin{equation}\label{eq:gausswkb}
\int_0^{x_+(\mi t_{N,k})}\sqrt{\me^{-2x^2}-t_{N,k}^2}\,\dif x=\frac{\sqrt{\pi}}{2N}\left(k+\frac{1}{2}\right),\;\;k=0,1,2,\ldots N-1,
\end{equation}
where $x_\pm$ are given by 
\begin{equation}\label{eq:gxplus}
x_\pm(\mi t)=\pm\sqrt{-\ln t}.
\end{equation}
Equation \eqref{eq:wkbeval} thus reduces in this case to  
\begin{equation}
\int_0^{\sqrt{-\ln t_{N,k}}}\sqrt{\me^{-2x^2}-t_{N,k}^2}\,\dif x=\frac{\sqrt{\pi}}{2 N}\left(k+\frac{1}{2}\right),\;\;k=0,1,2,\ldots N-1\,.
\label{eq:gwkbeval}
\end{equation}
It is the solutions $t_{N,k},\,k=0,\ldots,N-1$ of \eqref{eq:gwkbeval} together with the $\gamma_{N,k}$'s which will generate the exact $N$-soliton solution of \eqref{eq:nls}. The collection of these solutions for $N\in\NN$ is the \emph{Gaussian SSE}.

Our first task is to solve \eqref{eq:gwkbeval} to very high precision. With the numerically computed WKB spectral data in hand, we then use the numerical linear algebra routines of \cite{MK} and \cite{LM} to reconstruct via inverse scattering various members of the SSE at $t=0$ (and later times too). High precision knowledge of the spectral data is necessary due to fact that the solution is obtained by solving a poorly conditioned linear system \cite{MK}. We will then compare the numerical reconstructions of members of the Gaussian SSE at $t=0$ to the true initial data $A_0=\me^{-x^2}$ and with approximations to the evolution at later times. 

\begin{figure}[ht]
\centering
\begin{tabular}{cc}
\includegraphics[height=2.5in]{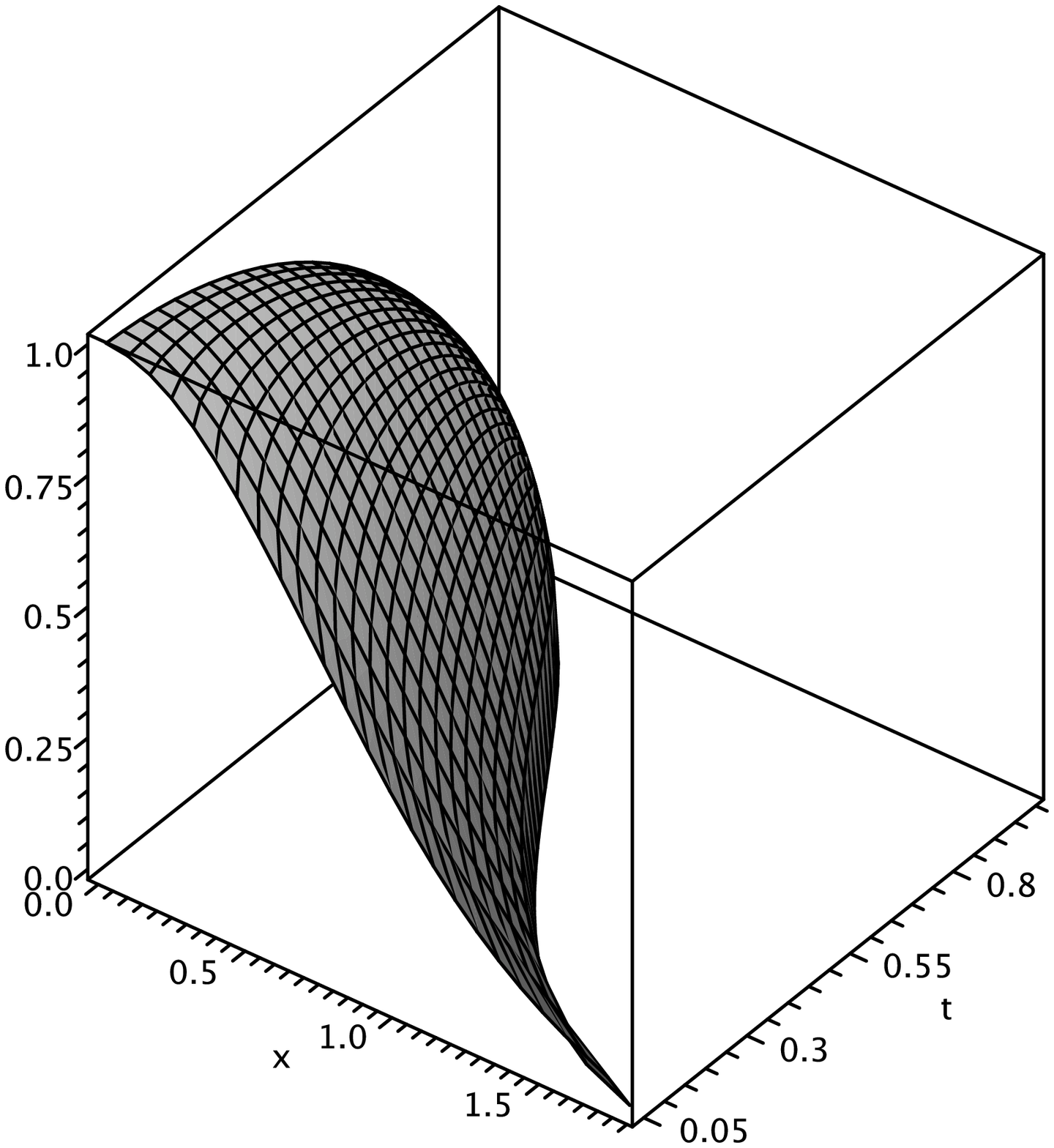} & 
\includegraphics[height=2.5in]{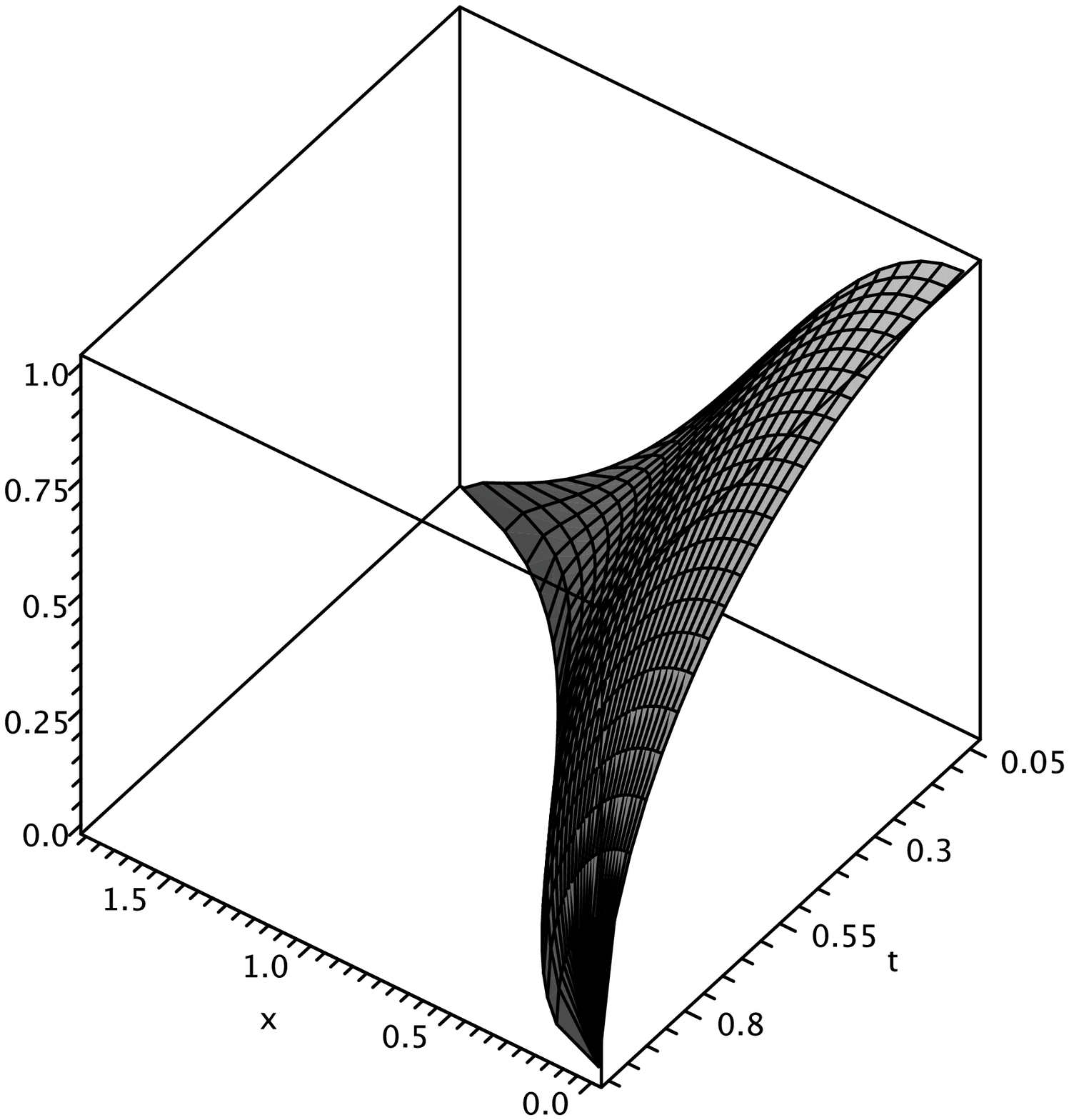}
\end{tabular}
\caption{Two pictures of the graph of $K$. Note that $K$ vanishes along the curve defined by  $(x_+(\mi t),t)=(\sqrt{-\ln t},t)$.} 
\label{fig:K}
\end{figure}

We now make a few comments about the solution of \eqref{eq:gwkbeval}. We performed these calculations with 250-digit precision in \textsc{Maple}. However, our initial attempts to solve equation \eqref{eq:gwkbeval} directly were unsuccessful, and we found it necessary to transform the problem to avoid difficulties with the root finder.   
In particular, if we define
\begin{equation}
F(t):=\int_0^{\sqrt{-\ln t}}\underbrace{\sqrt{\me^{-2x^2}-t^2}}_{K(x,t)}\,\dif x\,,
\label{eq:f}
\end{equation}
then equation \eqref{eq:gwkbeval} can be viewed as solving the single equation 
\[
F(t)=\text{constant}\,.
\]
Two views of the graph of the function $K$ appearing in the definition of $F$ in \eqref{eq:f} are shown in Figure~\ref{fig:K}, and we attributed the failure of the Newton solver to the square-root vanishing of $K$ and its influence on $F'(t)$ along the curve $(x_+(\mi t),t)$. 
To overcome the problem and eliminate the difficulty, we define 
\begin{equation}\label{eq:cov}
\me^{-2x^2}=t^2\cosh^2w\,,
\end{equation} and we write $F(t)$ as 
\begin{align}
F(t)&=\int_0^{\frac{1}{2}\sqrt{-\ln t}}\sqrt{\me^{2x^2}-t^2}\,\dif x + \int_{\frac{1}{2}\sqrt{-\ln t}}^{\sqrt{-\ln t}}\sqrt{\me^{2x^2}-t^2}\,\dif x \\
&=: F_\mathrm{I}(t)+F_\mathrm{II}(t)\,. \nonumber
\end{align}
Then, changing the integral in $F_\mathrm{II}$ via \eqref{eq:cov} to an integral with respect to $w$, we obtain
\begin{align}\label{eq:f2new}
F_\mathrm{II}(t) &=\int_{\frac{1}{2}\sqrt{-\ln t}}^{\sqrt{-\ln t}}\sqrt{\me^{2x^2}-t^2}\,\dif x 
	=\frac{1}{2}\int_0^{\cosh^{-1}(t^{-3/4})}\frac{t\sinh w\tanh w}{\sqrt{-\frac{1}{2}\ln(t^2\cosh^2 w)}}\,\dif w\,. 
\end{align}
It follows that \eqref{eq:gwkbeval} may be rewritten as 
\begin{multline}
\int_0^{\frac{1}{2}\sqrt{-\ln t_{N,k}}}\sqrt{\me^{-2x^2}-t_{N,k}^2}\,\dif x+\frac{1}{2}\int_0^{\cosh^{-1}(t_{N,k}^{-3/4})}\frac{t_{N,k}\sinh w\tanh w}{\sqrt{-\frac{1}{2}\ln(t_{N,k}^2\cosh^2 w)}}\,\dif w\\ =\frac{\sqrt{\pi}}{2N}\left(k+\frac{1}{2}\right)\,.\label{eq:wkbnew}
\end{multline}
This is the equation we solve to high precision. We verified the 250-digit accuracy of the solutions of \eqref{eq:wkbnew} using both \textsc{Mathematica} and \textsc{Maple} routines.

\section{Numerical Methods}\label{sec:methods}

We introduce an implicit finite difference scheme to solve the initial value problem (\ref{eq:nls_ivp}) in this section. 
In Section \ref{sec:experiments} we use an exact solution to illustrate the order of accuracy of the proposed method. We show that the temporal grid sizes used for the proposed method scale linearly with the spatial mesh refinement. We then use particular $N$-soliton solutions, members of the Satsuma--Yakima ensemble \cite{SY}, which we obtain by the IST calculation, to validate the proposed method for small-$\epsilon$ calculation. At the same time, we compare the proposed method with the well-known spectral split-step method and show that the proposed method is a suitable method for solving the focusing NLS in the semiclassical regime. We also investigate a filtering process that removes Fourier modes whose amplitudes are smaller than a given threshold for our calculations with small $\epsilon$. Finally, in Section \ref{sec:gauss}, we compare numerical solutions of the proposed method with that of the Gaussian SSE for the focusing NLS.
 
The focusing NLS equation (\ref{eq:nls}) we consider here can be easily rescaled into the standard cubic NLS 
\begin{equation}\label{eq:cubic_nls}
\mi\pd_{t^*} \psi+\pd_{x^*}^2\psi+2 |\psi|^2\psi=0\,,
\end{equation}
with $u(t,x)=\sqrt{\epsilon}\psi(2t^*,\sqrt{\epsilon} x^*)$. Equation (\ref{eq:cubic_nls}) is completely integrable in the sense of IST and has a canonical Hamiltonian form. A spatial finite-difference semi-discretization of equation (\ref{eq:cubic_nls}) proposed by Ablowitz and Ladik \cite{AL},
\beq\label{eq:AL_nls}
\mi\frac{\dif\psi_m}{\dif t} +\frac{1}{h^2}\left(\psi_{m+1}-2\psi_{m}+\psi_{m-1}\right) + |\psi_m|^2\left(\psi_{m+1}+\psi_{m-1}\right)=0\,,
\eeq
is also completely integrable and posseses a Hamiltonian structure; here, $h$ is the spatial grid size. We refer to the above discretization as the AL-lattice. Fornberg \cite{Fornberg} has shown that with accurate (exact) time integration, the AL-lattice is very suitable for numerical work, since it produces few numerical artifacts for unstable analytical solutions in a periodic domain. More discussion on numerical homoclinic instability for the standard NLS can be seen, for example, in the papers by Ablowitz et al. \cite{AH, AHC}.  Nevertheless, choosing a proper numerical time integrator for the AL-lattice is by no means a trivial task. 


Schober et al. \cite{Schober, IKS} indicate that the Hamiltonian system of the AL-lattice carries on its phase space a noncanonical symplectic structure for which standard symplectic integrators, such as symplectic implicit Runge--Kutta methods, are not immediately applicable.
 Several approaches are provided by Schober et al. \cite{Schober, IKS} to remedy the situation. While symplectic algorithms have the advantage of preservation of the global and local conservation laws for a long period of time, our aim is in the direction of developing an efficient and stable algorithm that is suitable for fine-grid calculations in order to accurately capture the behavior of \eqref{eq:nls} in the small-$\epsilon$ regime. 
 
Taking advantage of the simple and clean form of the AL-lattice, we propose to apply the implicit midpoint time integrator to the AL-lattice directly. The implicit midpoint method is the lowest order member of the Gauss--Legendre family of implicit Runge--Kutta methods which are symplectic schemes for canonical Hamiltonian systems \cite{IKS}. Our numerical experiments show that the combination of the midpoint time integrator and the AL-lattice is advantageous for solving the semiclassical focusing NLS equation. The advantages include (1) the ratio of temporal grid size and the spatial grid size used for the method can be kept constant when refining the mesh; (2) with a good initial guess, the simple fixed-point iteration process converges relatively fast ($<10$ iterations for the convergence tolerance $\gamma \le 10^{-12}$); (3) unlike the standard spectral split-step method, the proposed method is less sensitive to what spatial and temporal grid sizes to use in the simulations to avoid numerical artifacts caused by numerical roundoff error for small $\epsilon$.  

Finally, we remark that many numerical methods for the focusing NLS in the semiclassical regime have been discussed in the literature \cite{BK, BJM1,  BJM, CT}, but none directly compared with the IST calculation.  

\subsection{Implicit Finite Difference Algorithm} The proposed finite difference scheme for the initial value problem (\ref{eq:nls_ivp}) is as follows.

\vspace{12pt}

\noindent
{\bf Step 1.} Based on the AL-lattice, the spatial discretization of equation (\ref{eq:nls}) is
\beq\label{eq:AL_nls_focusing}
\mi\epsilon\frac{\dif u_m}{\dif t} +\frac{\epsilon^2}{2\Delta x^2}\left(u_{m+1}-2u_{m}+u_{m-1}\right) +\frac{1}{2} |u_m|^2\left(u_{m+1}+u_{m-1}\right)=0\,,
\eeq
where $\Delta x$ is the spatial grid size, and $u_m$ approximates the solution at the $m^{th}$ grid point.

\vspace {12pt}

\noindent
{\bf Step 2.} Applying the midpoint time integrator to the above ordinary differential equations (ODE) yields
 \beq\label{eq:AL_midpoint}
 u_m^{n+1} = u_m^{n} +\frac{\mi\epsilon \Delta t}{2\Delta x^2}\left(u_{m+1}^{n+1/2}-2u_m^{n+1/2}+u_{m-1}^{n+1/2}\right) +\frac{\mi \Delta t}{2\epsilon}|u_m^{n+1/2}|^2\left(u_{m+1}^{n+1/2}+u_{m-1}^{n+1/2}\right),
 \eeq
where $\Delta t$ is the time step size, and $u_m^{n+1/2}$ is defined as
\beq\label{eq:half_step1}
u_m^{n+1/2} = \frac{1}{2}\left(u_m^{n}+u_m^{n+1}\right)\,.
\eeq 

\vspace {12pt}

\noindent
{\bf Step 3.}  For $n=1\dots N$,  we solve the nonlinear equations (\ref{eq:AL_midpoint}) by using the simple fixed-point-iteration (FPI) procedure, in which the $(k+1)^{th}$ iteration is written as
 \beq\label{eq:FPI}
 \begin{split}
 u_m^{n+1, (k+1)} = & u_m^{n} +\frac{\mi\epsilon \Delta t}{2\Delta x^2}\left(u_{m+1}^{n+1/2, (k)}-2u_m^{n+1/2, (k)}+u_{m-1}^{n+1/2, (k)}\right)\\  & + \frac{\mi \Delta t }{2\epsilon}|u_m^{n+1/2, (k)}|^2\left(u_{m+1}^{n+1/2, (k)}+u_{m-1}^{n+1/2, (k)}\right)\,,
 \end{split}
 \eeq
where $u_m^{n+1/2, (k)}$ is defined as
\beq\label{eq:half2}
u_m^{n+1/2, (k)} = \frac{1}{2}\left(u_m^{n}+u_m^{n+1,(k)}\right).
\eeq
The initial guess $u_m^{n+1,(0)}$ for the FPI procedure within each time step is 
the solution of the Crank-Nicolson-type scheme for the AL-lattice:
\beq\label{eq:AL_CN}
 u_m^{n+1} = u_m^{n} +\frac{\mi\epsilon \Delta t}{2\Delta x^2}\left(u_{m+1}^{n+1/2}-2u_m^{n+1/2}+u_{m-1}^{n+1/2}\right) +\frac{\mi \Delta t}{2\epsilon}|u_m^{n}|^2\left(u_{m+1}^{n}+u_{m-1}^{n}\right)\,,
 \eeq
where $u_m^{n+1/2}$ is defined in (\ref{eq:half2}). Equation (\ref{eq:AL_CN}) results in a tridiagonal system for $u_m^{n+1}$, which is solved by the Thomas Algorithm \cite{KK}. The convergence tolerance  for the FPI procedure is 
\beq
\|u^{n+1, (k+1)}-u^{n+1, (k)}\|_{\infty} \le \gamma\,,
\eeq
where $\gamma\le10^{-12}$ for the numerical experiments throughout this paper. Here $\|\cdot\|_{\infty}$ is the infinity-norm defined by 
\beq\label{eq:inf_norm}
\|u\|_{\infty}=\underset{m=1,\dots M}{\text{max}}\,\,|u_m|\,.
\eeq
When the convergence tolerance  is achieved, we set $u^{n+1}=u^{n+1, (k+1)}$, and move onto the next time step. 


\subsection{Spectral Split-Step Method}\label{sec:SS}
Splitting schemes are very appealing for solving the focusing NLS equation in periodic domains. Within one $\Delta t$, a splitting method advances the NLS equation (\ref{eq:nls}) by solving the following two equations alternately. 

\vspace{12pt}
 
\noindent
(A) Nonlinear part (solve exactly in physical space)
\beq\label{eq:ODE1}
u_t=\frac{2\mi}{\eps}|u|^2u\,.
\eeq
(B) Linear part (solve exactly in Fourier space):
\beq\label{eq:ODE2}
u_t=\mi\eps u_{xx}\,.
\eeq
Yoshida \cite{YH}  introduced a systematic method to construct arbitrary even-order time accurate splitting schemes. For example, to obtain second-order accuracy in time, we solve the two equations sequentially, like (A) $\longrightarrow$ (B) $\longrightarrow$  (A), by using the time increments $\{\frac{\Delta t}{4},\,\frac{\Delta t}{2},\,\frac{\Delta t}{4}\}$ in each step, respectively. Alternatively, one can also solve the sequence (B) $\longrightarrow$ (A) $\longrightarrow$  (B) to obtain the same order of accuracy, although this sequence is more time consuming, since one has to compute the (inverse) Fast Fourier Transform twice. We remark that the second-order accurate method constructed by using the Yoshida's scheme is essentially the Strang splitting method \cite{BJM}. The sequences and time increments for fourth and sixth-order methods are listed, for example, in the paper by Fornberg \& Driscoll \cite{FD}.  

In Step (A), we solve the ODE (\ref{eq:ODE1}) exactly. Bao et al. \cite{BJM} did a simple calculation to show that  $|u|^2$ in equation (\ref{eq:ODE1}) is invariant within each time increment,
\beq\label{eq:invariant}
\partial_t |u|^2 = 2\text{Re}(u_t\,\bar{u})
                       = \frac{4}{\epsilon}\text{Re}(\mi |u|^2u\bar{u}) 
                       = \frac{4}{\epsilon}\text{Re}(\mi |u|^4)
                       =  0\,.
\eeq
Equation (\ref{eq:invariant}) implies that the ODE (\ref{eq:ODE1}) is linear and separable.

\vspace{12pt}

\noindent
We use the second-order scheme to illustrate the split step algorithm as follows. Suppose that in a periodic domain, equation (\ref{eq:nls}) is solved on a mesh for $u_m$, where $m=1\dots M+1$  and $u_{M+1}=u_{1}$, within a time step $\Delta t=t^{n+1}-t^{n}$. 

\noindent
{\bf Step 1:} We solve (A) in the physical space with the initial time $t^n$ and the final time  $t^{n}+\frac{1}{4}\Delta t$. The solution at the $m^{th}$ grid point is
\beq\label{eq:sol1}
u_m(t^n+\frac{1}{4}\Delta t) =u_m(t^n)e^{\frac{\mi\Delta t}{2\epsilon}|u_m(t^n)|^2}\,. 
\eeq

\noindent
{\bf Step 2:} We solve (B) in the Fourier space. The Fourier transform of equation (\ref{eq:ODE2}) is
\beq\label{eq:ODE2_FT}
\hat{u}_t=-\mi k^2 \epsilon \hat u\,,
\eeq
where $k$ is the wavenumber, and $\hat{u}$ is the Fourier transform of $u$. The Fourier transform pair for $u$ are defined as
\beq\label{eq:FFT}
\begin{split}
\hat{u}_k&=\sum_{m=1}^{M}u_m\omega_M^{(m-1)(k-1)}\,,\\
u_m&=\frac{1}{M}\sum_{k=1}^{M}\hat{u}_{k}\omega_{M}^{-(m-1)(k-1)}\,,
\end{split}
\eeq
where
\beq
\omega_M=e^{(-2\pi\mi)/M}\,.
\eeq
Solving the ODE (\ref{eq:ODE2_FT}) with the initial time $t^{n}+\frac{1}{4}\Delta t$ and the final time $t^{n}+\frac{3}{4}\Delta t$ yields the 
$k^{th}$ Fourier mode of $u$,
\beq\label{eq:sol2}
\hat{u}_k(t^n+\frac{3}{4}\Delta t) =\hat{u}_k(t^{n}+\frac{1}{4}\Delta t)e^{-\frac{1}{2}\mi k^2 \epsilon\Delta t}\,. 
\eeq

\vspace{6pt}

\noindent
{\bf Step 3:} After taking the inverse Fourier transform of $\hat{u}$ in (\ref{eq:sol2}), we solve (A) again in the physical space with the initial time $t^n+\frac{3}{4}\Delta t$ and the final time  $t^{n+1}$. The solution at the $m^{th}$ grid point is
\beq
u_m(t^{n+1}) =u_m(t^n+\frac{3}{4}\Delta t)e^{\frac{\mi\Delta t}{2\epsilon}|u_m(t^n+\frac{3}{4}\Delta t)|^2}\,. 
\eeq

\section{Numerical Experiments}\label{sec:experiments}

\subsection{Exact Solution}

It is easy to check that an exact solution associated with the focusing NLS equation (\ref{eq:nls}) is
\beq\label{eq:exact}
u(x,t)=\sech\left(\frac{x+2t}{\epsilon}\right)\me^{\frac{-2\mi}{\epsilon}(x+\frac{3}{4} t)}\,.
\eeq
The structure of the solution \eqref{eq:exact} is simple; it features a bell-shaped envelope propagating at a constant rate.
We use this exact solution to validate our numerical implementation and to test the order of accuracy of the numerical methods. Table \ref{tab:order} is the mesh refinement study of the proposed implicit finite difference method and the split-step method. In the calculations, the length of the periodic domain is $[-16,\,16]$, the small parameter is $\epsilon=0.5$, and the final time is $t=0.5$. The temporal grid size for the finite difference method is $\Delta t/\Delta x=1/80$, and the convergence tolerance is $\gamma=10^{-13}$. If we define the physical quantity
\begin{equation}\label{eq:u2}
\rho(x,t)=|u(x,t)|^2\,,
\end{equation}
the error of $\rho$ between the numerical calculation and the exact solution measured by the 2-norm 
\beq\label{eq:2norm}
\|\rho\|_2=\sqrt{\frac{1}{M}\sum_{m=1}^{M} \rho_m^2}\,,
\eeq
is shown in Table \ref{tab:order}. The table indicates that both methods are second-order accurate. We also note that for both methods, $\Delta t/\Delta x$ can be kept roughly constant throughout the mesh refinement study to maintain the desired accuracy.  

\begin{table}[htpb]
\caption{Mesh refinement study}
\label{tab:order}
\centering
\begin{tabular}{c|cccccc}
\hline
\backslashbox{}{$\Delta x$} & 1/32 &  1/64 & 1/128 & 1/256 & 1/512 & 1/1024\\\hline
$\|\rho_{\text{FD}}-\rho_{\text{exact}}\|_2$ & 5.0836e-04 & 1.2709e-04 & 3.1772e-05 & 7.9429e-6 & 1.9857e-6 & 4.9643e-07\\\hline 
$\Delta t/\Delta x$ & 1/80 & 1/80 & 1/80 & 1/80 & 1/80 & 1/80\\\hline
order & & 2 & 2 & 2 & 2 & 2\\\hline
$\|\rho_{\text{SS}}-\rho_{\text{exact}}\|_2$ & 1.1046e-06 & 2.7624e-07 & 6.8953e-08 & 1.7372e-8 & 4.4907e-9 & 1.1117e-09\\\hline
$\Delta t/\Delta x$ & 1/10 & 1/10 & 1/10 & 1/10 & 1/10 & 1/13\\\hline
order & & 2 & 2 & 1.99 & 1.95 & 2.01\\\hline
\end{tabular}
\end{table}

\subsection{$N$-soliton} Lyng \& Miller \cite{LM} introduced accurate numerical reconstructions of the $N$-soliton by the IST; these refined earlier calculations of  Miller \& Kamvissis \cite{MK}. In this case, the $N$-soliton is the solution of the initial-value problem 
\beq\label{N-soliton}
\begin{split}
\mi\psi_t+\frac{1}{2}\psi_{xx}+|\psi|^2\psi = 0\,, \\
\psi(x,0)=N\sech(x)\,.
\end{split}
\eeq
If the amplitude and the time variable of $\psi$ are scaled by a parameter $\epsilon$ for a new variable, $u(x,t)=\epsilon\psi(x, t/\epsilon )$, the initial value problem (\ref{N-soliton}) is equivalent to the focusing NLS equation for $u$:
\beq\label{N-soliton-e}
\begin{split}
\mi\epsilon u_t+\frac{1}{2}\epsilon^2u_{xx}+|u|^2u = 0\,, \\
u(x,0)=A\sech(x)\,,
\end{split}
\eeq
where $\epsilon=A/N$. It is well-known that the $N$-soliton breaks its focusing state into the oscillatory state at the critical time $t_c=(2A)^{-1}$ \cite{KMM}. We consider the case of initial amplitude $A=2$ (thus $t_c=0.25$) and compute the initial value problem (\ref{N-soliton-e}) for various $N$ by using the IST. We choose the time slice at $t=0.3$ ($>t_c$) with $4096$ points in the interval $0\le x\le 1$ on the $(t,\,x)$ plane. All calculations are done using \textsc{Mathematica} with 250-digit precision. The solutions obtained by the IST calculation are, up to the numerical precision of the implementation of the IST, exact solutions of the $N$-soliton problem \eqref{N-soliton-e}. We use these solutions to assess the performance of the finite difference and the spectral split-step algorithms for small $\epsilon$. 

We now test these two numerical algorithms for the $N$-soliton problem with the initial data $u(x,0) = 2\sech(x)$. The periodic computational domain is $-16\le x\le 16 $. We note that results obtained by using the initial data $u(x,0) = 2\sech(x)$ ($\epsilon=2/N$) at the final time $t=t_f$ are equivalent to those that computed with initial data $u(x,0)=\sech(x)$ ($\epsilon=1/N$) at the final time $t=2 t_f$, modulo a factor two in amplitude.\\

\noindent
{\bf Example 1:} 
$
u(x,0) = 2\sech(x),\,\, N=40,\,\, \epsilon=0.05,\,\,\text{final time}\,\, t=0.3.\\
$

\begin{figure}[hbtp]
\centering
(a)\includegraphics[width=3in]{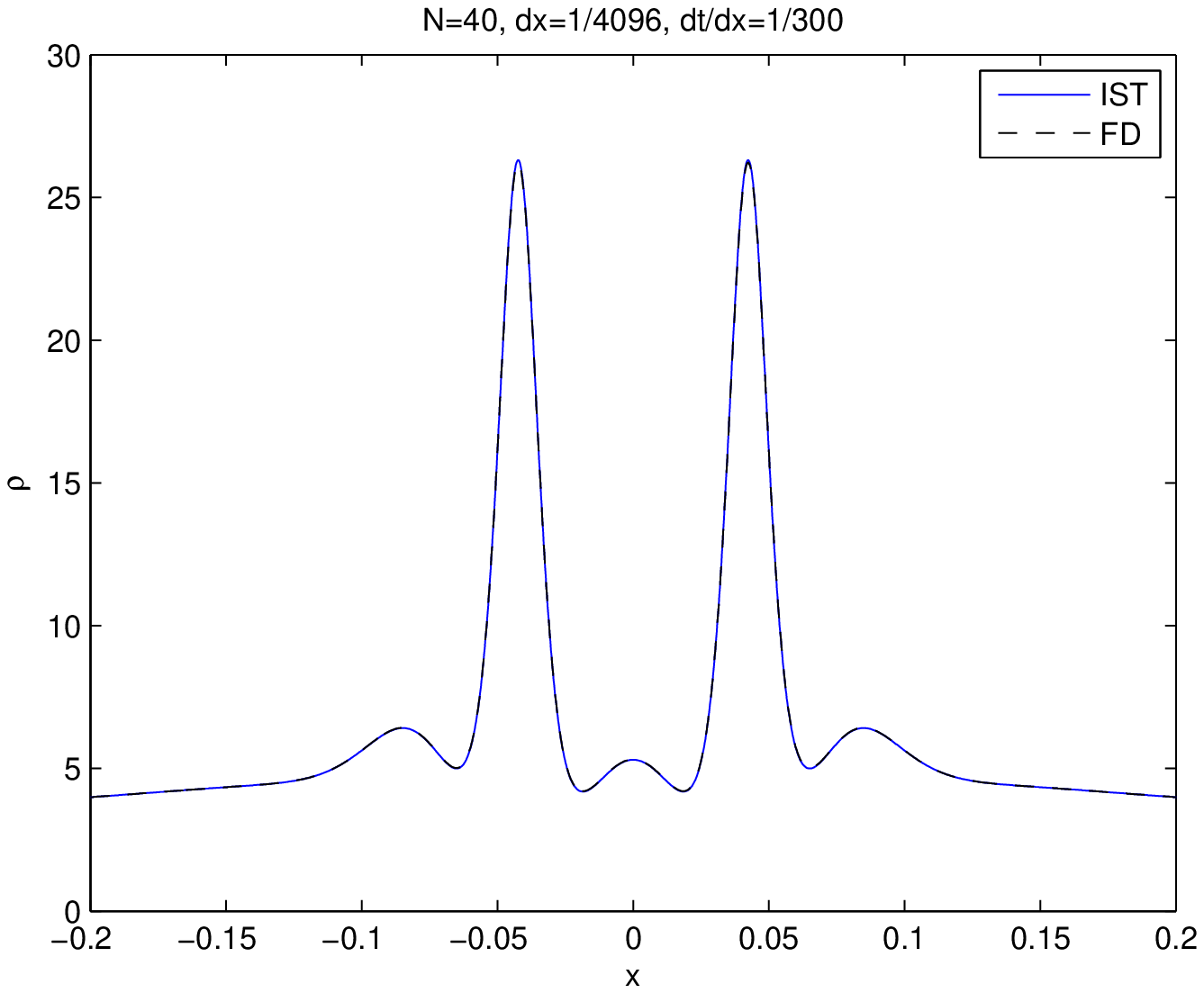}
(b)\includegraphics[width=3in]{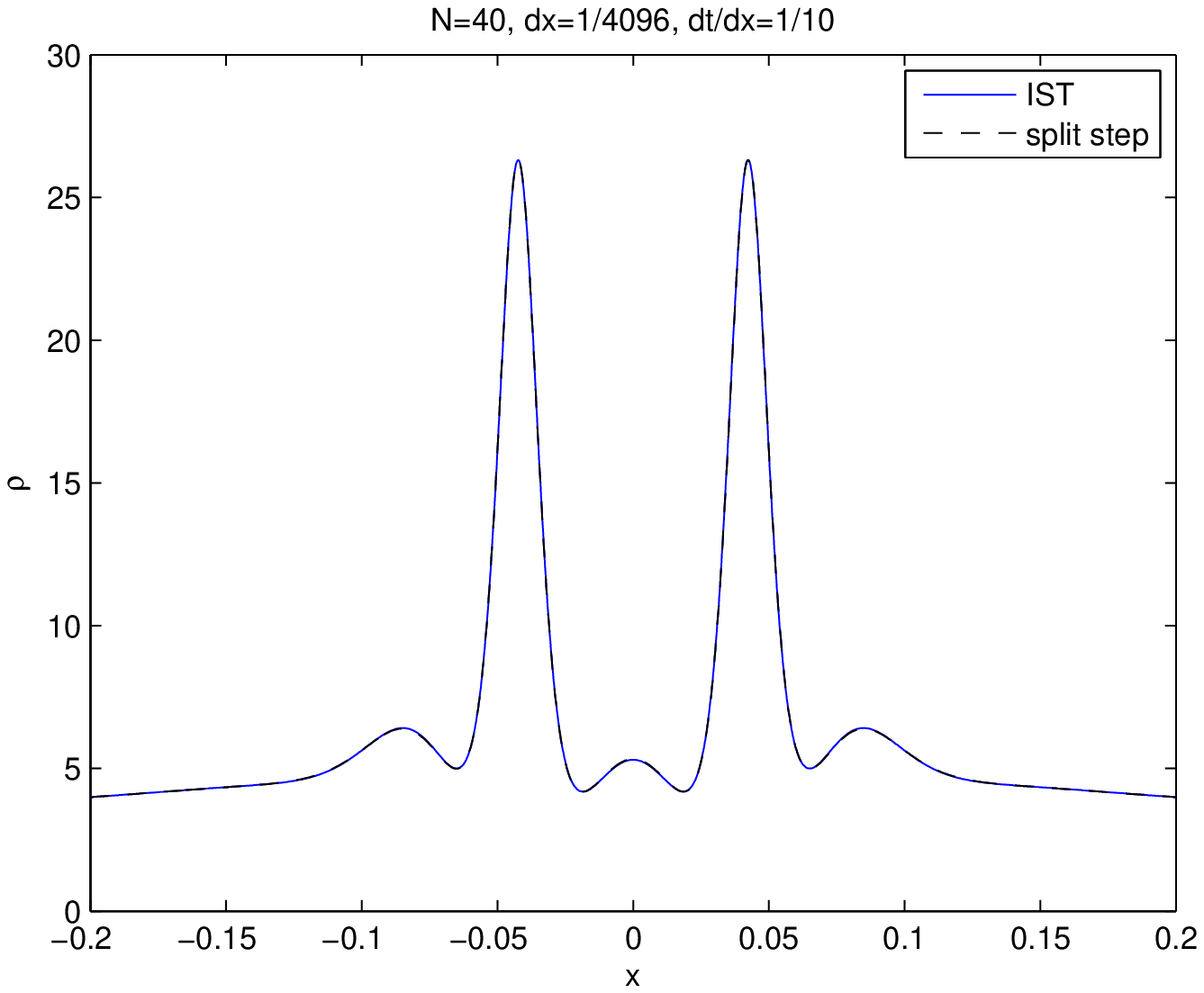}
\caption{$N=40$ at $t=0.3$. (a) A comparison of the finite difference method and the IST. The $2$-norm error is $9\times 10^{-3}$.  (b) A comparison of the spectral split-step method and the IST. The $2$-norm error is $7.5\times 10^{-3}$.}
\label{fig:N40_t03}
\end{figure}

\noindent
Figure \ref{fig:N40_t03} shows simulations for $N=40$ at the final time $t=0.3$.  Figure \ref{fig:N40_t03}(a) plots the conserved quantity $\rho$ computed by using the finite difference method. Figure \ref{fig:N40_t03}(b) is the counterpart of  Figure \ref{fig:N40_t03}(a), using the spectral split-step method. Both (a) and (b) are plotted against the IST solution. We observe that numerical solutions of both methods are visually indistinguishable from the quasi-exact solution. For the finite difference method, the 2-norm error against the IST solution is $9\times 10^{-3}$, and for the spectral split-step method, the error is $7.5\times 10^{-3}$. All 2-norm errors in this section are measured for $-1\le x\le 1$, unless specified otherwise. The finite difference method uses the spatial grid size $\Delta x=1/4096$ and the temporal grid size $\Delta t/\Delta x=1/300$. The convergence tolerance is $\gamma=10^{-12}$. The spectral split-step method uses $\Delta x=1/4096$ and $\Delta t/\Delta x=1/10$. Similar to the example with the exact solution \eqref{eq:exact}, this experiment indicates that the spectral split-step method is able to capture the right solution, and is more efficient than the finite difference method for $\epsilon=0.05$.  We note that similar output to Figure \ref{fig:N40_t03} can be obtained by using $\epsilon=0.025$ with the initial data $u(0,x)=\sech(x)$ at the final time $t=0.6$. We choose $A=2$ for a shorter breaking time.\\

\noindent
{\bf Example 2:} 
$
u(x,0) = 2\sech(x),\,\, N=50,\,\, \epsilon=0.04,\,\,\text{final time}\,\, t=0.3.\\
$

\begin{figure}[hbtp]
\centering
(a)\includegraphics[width=3in]{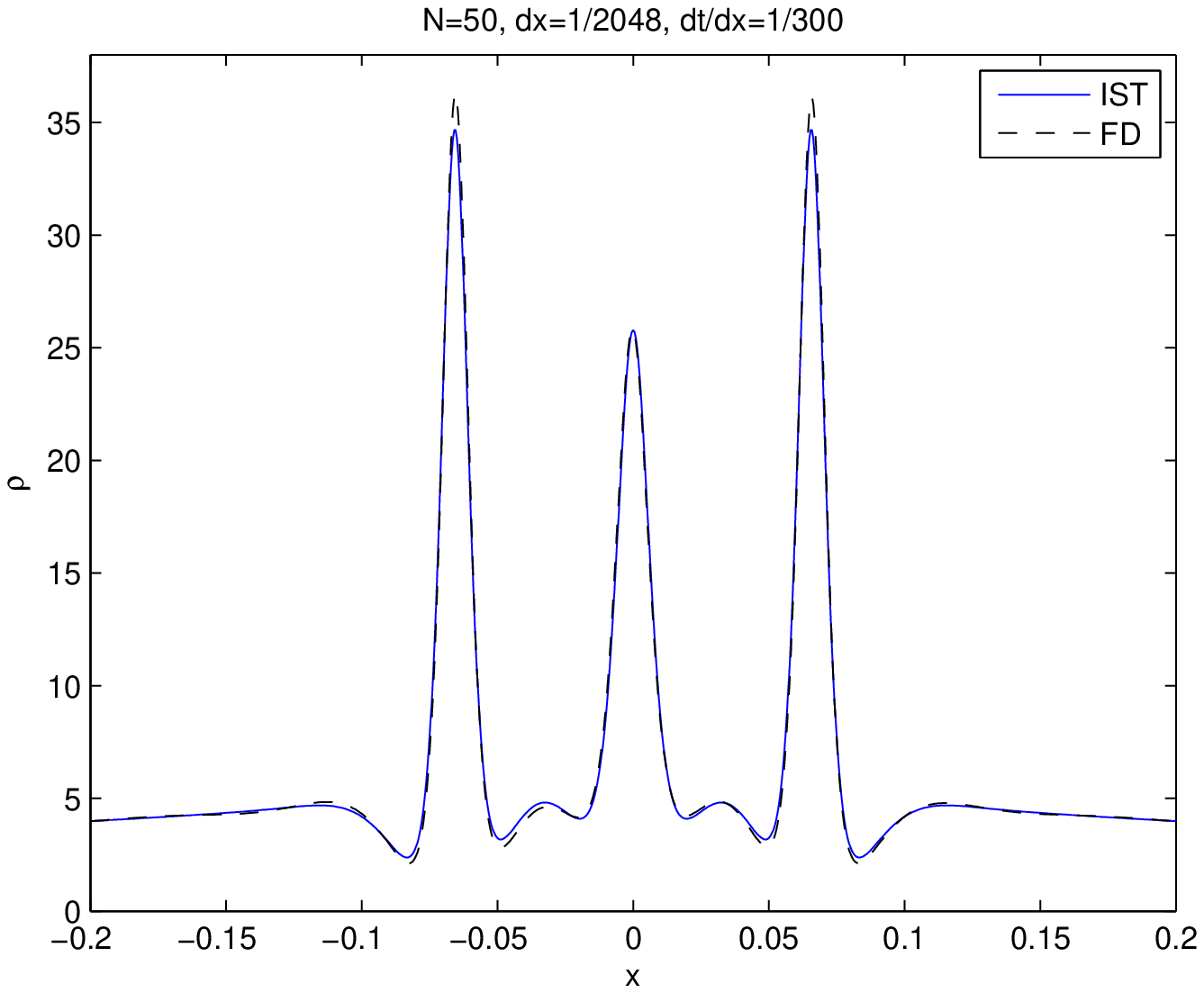}
(b)\includegraphics[width=3in]{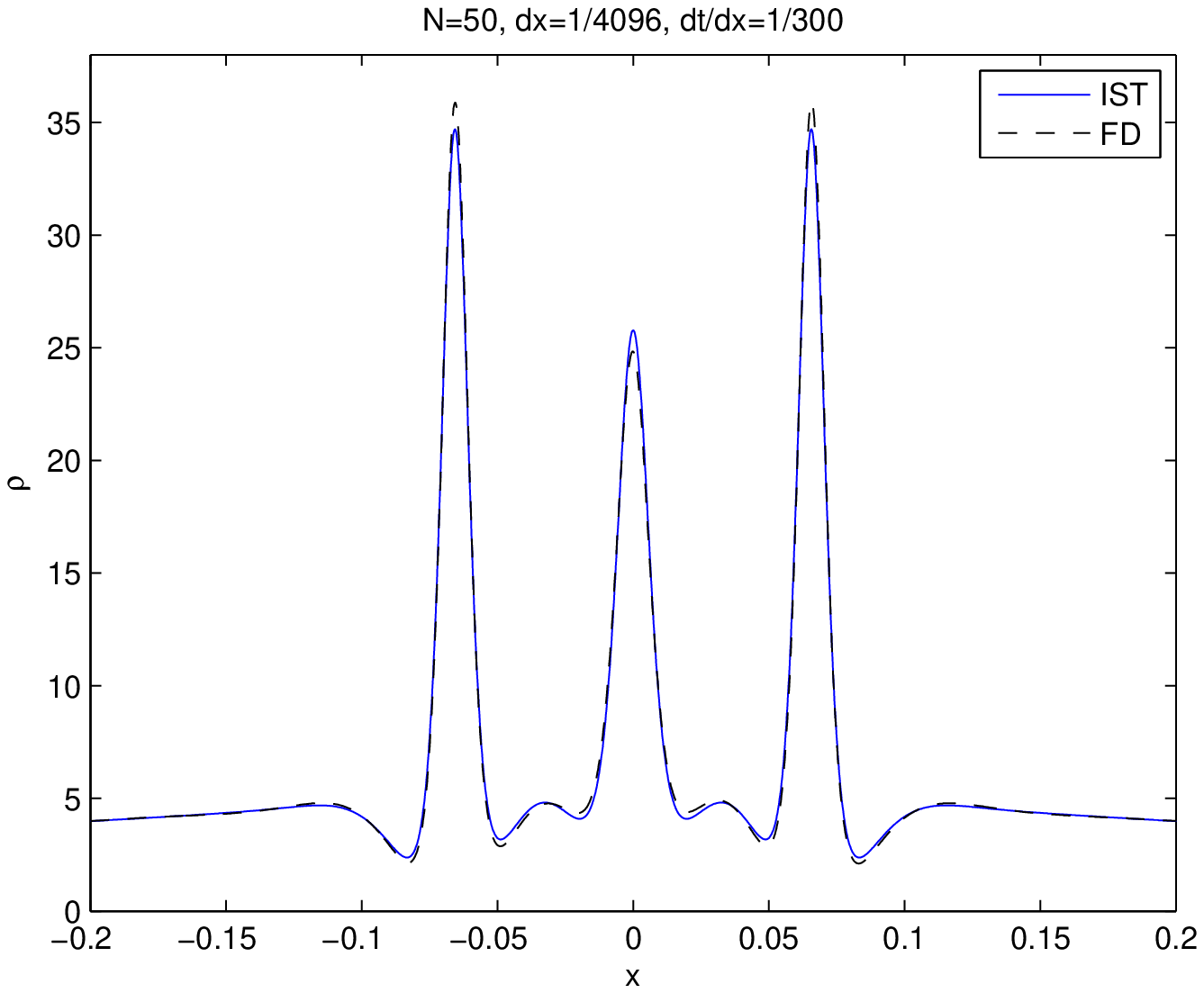}
\caption{$N=50$ at $t=0.3$. Refinement study of the finite difference method. 
(a)  $\Delta x=1/2048$, $\Delta t/\Delta x=1/300$, and the $2$-norm error is 0.1519. (b) $\Delta x=1/4096$, $\Delta t/\Delta x=1/300$, and the $2$-norm error is 0.1300.}
\label{fig:N50_FD_t03}
\end{figure}

\noindent
Figure \ref{fig:N50_FD_t03} shows a refinement study of the finite difference method for the $2\sech(x)$ initial data with $N=50$. The figure plots the computed quantity $\rho$ against the IST results.
%
%
Figure \ref{fig:N50_FD_t03}(a) uses $\Delta x=1/2048$ and $\Delta t/\Delta x=1/300$, and Figure \ref{fig:N50_FD_t03}(b) uses  $\Delta x=1/4096$ and $\Delta t/\Delta x=1/300$. We observe that the $2$-norm errors for (a) and (b) are $0.1519$ and $0.1300$, respectively. The convergence tolerances are $\gamma=10^{-15}$ for $\Delta x=1/2048$ and $\gamma=10^{-12}$ for $\Delta x=1/4096$. The study suggests that for the finite difference method, combined spatial and temporal refinement will reduce the error for $\epsilon=0.04$. We now compare  Figure \ref{fig:N50_FD_t03}(b) with the previous example, Figure \ref{fig:N40_t03}(a). We observe that while these two calculations use the same spatial and temporal discretization, the error in Figure \ref{fig:N50_FD_t03}(b) is one order larger than that in Figure \ref{fig:N40_t03}(a). The only difference between these two simulations is the small parameter $\epsilon$, for which one is $0.05$ and the other is $0.04$. This seems to indicate that when the focusing NLS equation becomes just a little bit more singular (i.e. $\epsilon$ decreases from 0.05 to 0.04), the roundoff error sets in rather swiftly. The phenomenon of roundoff error becomes much more prominent for the spectral split-step method.

\begin{figure}[hbtp]
\centering
(a)\includegraphics[width=3.in]{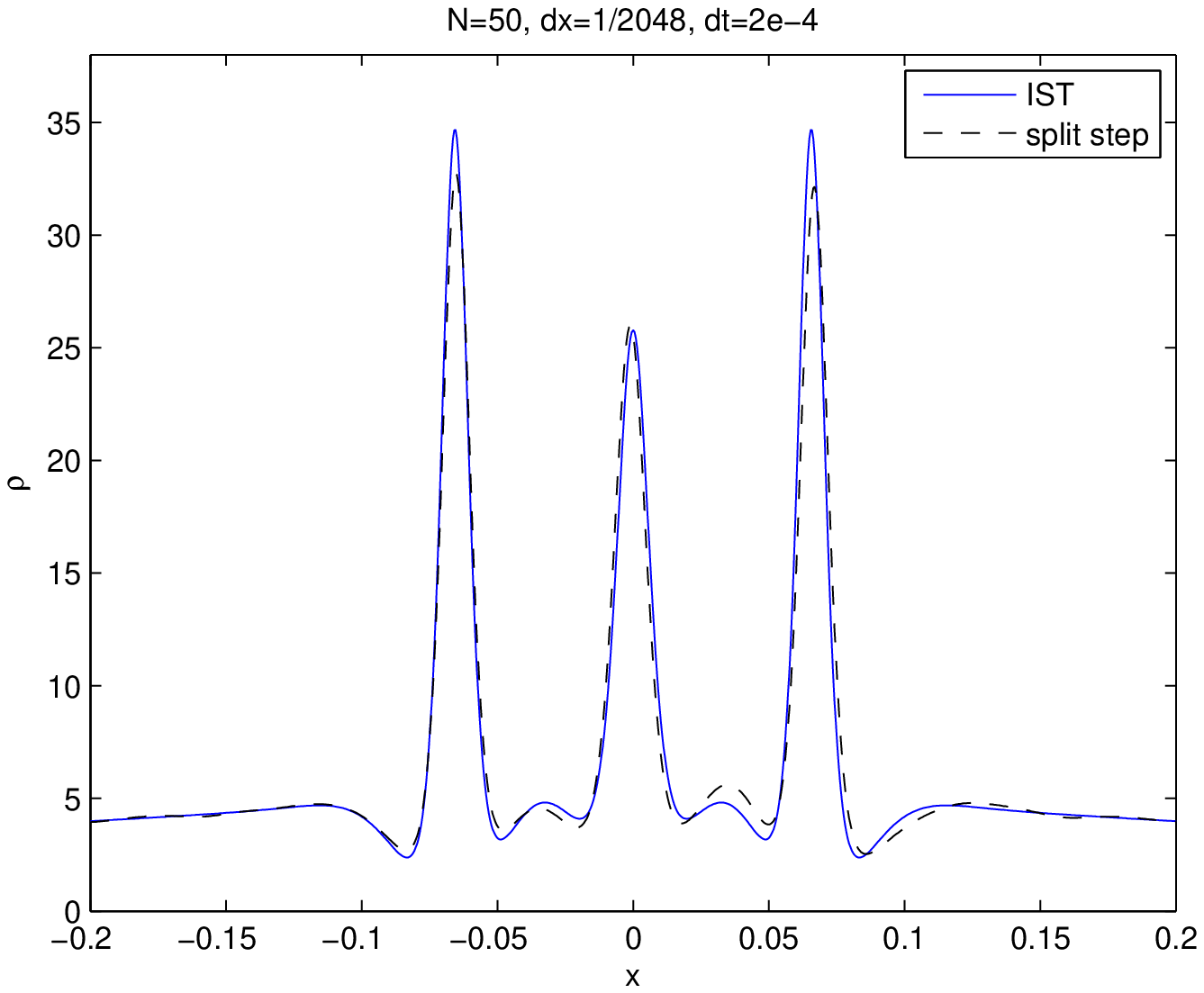}
\includegraphics[width=3.in]{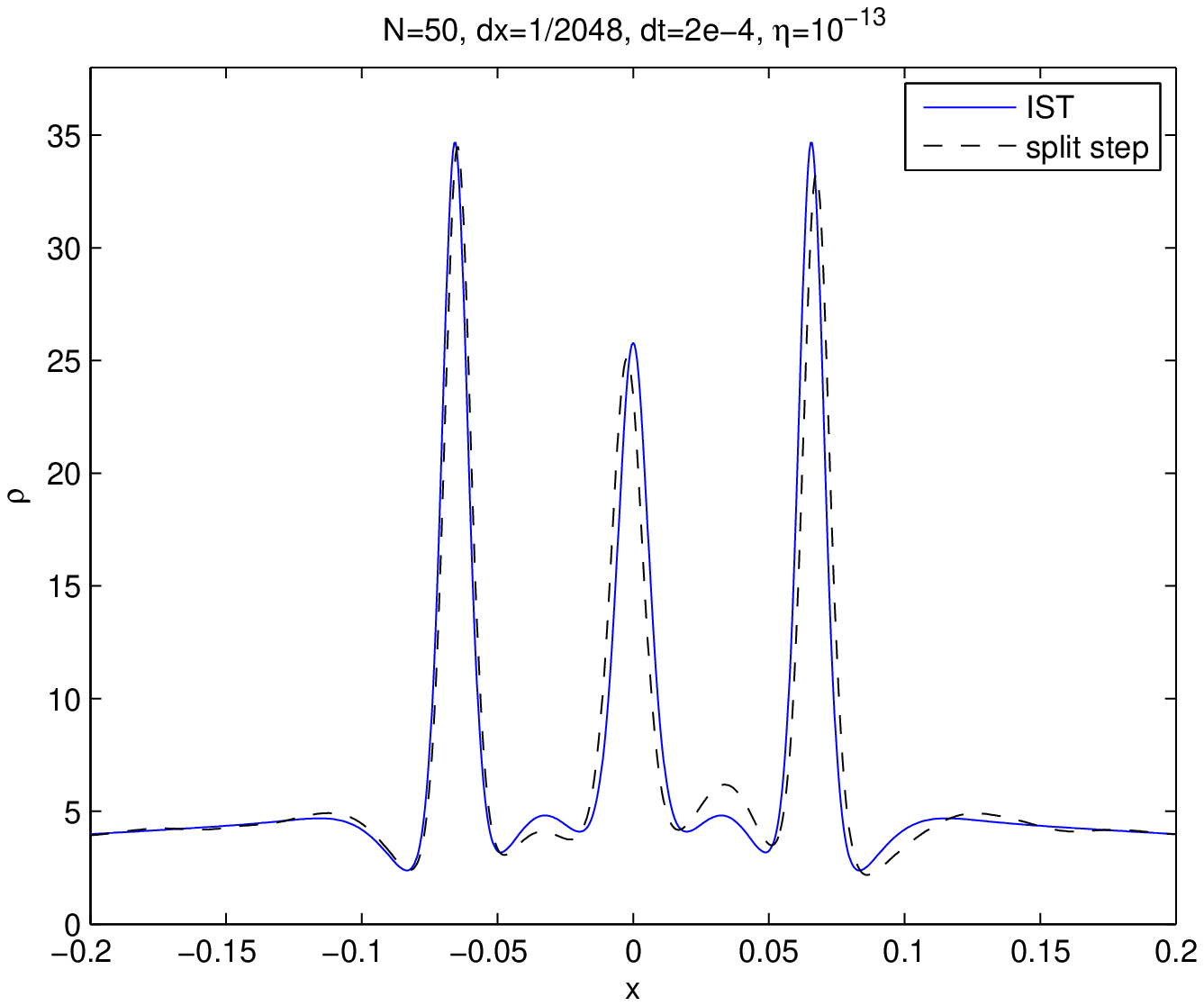}\\
(b)\includegraphics[width=3.in]{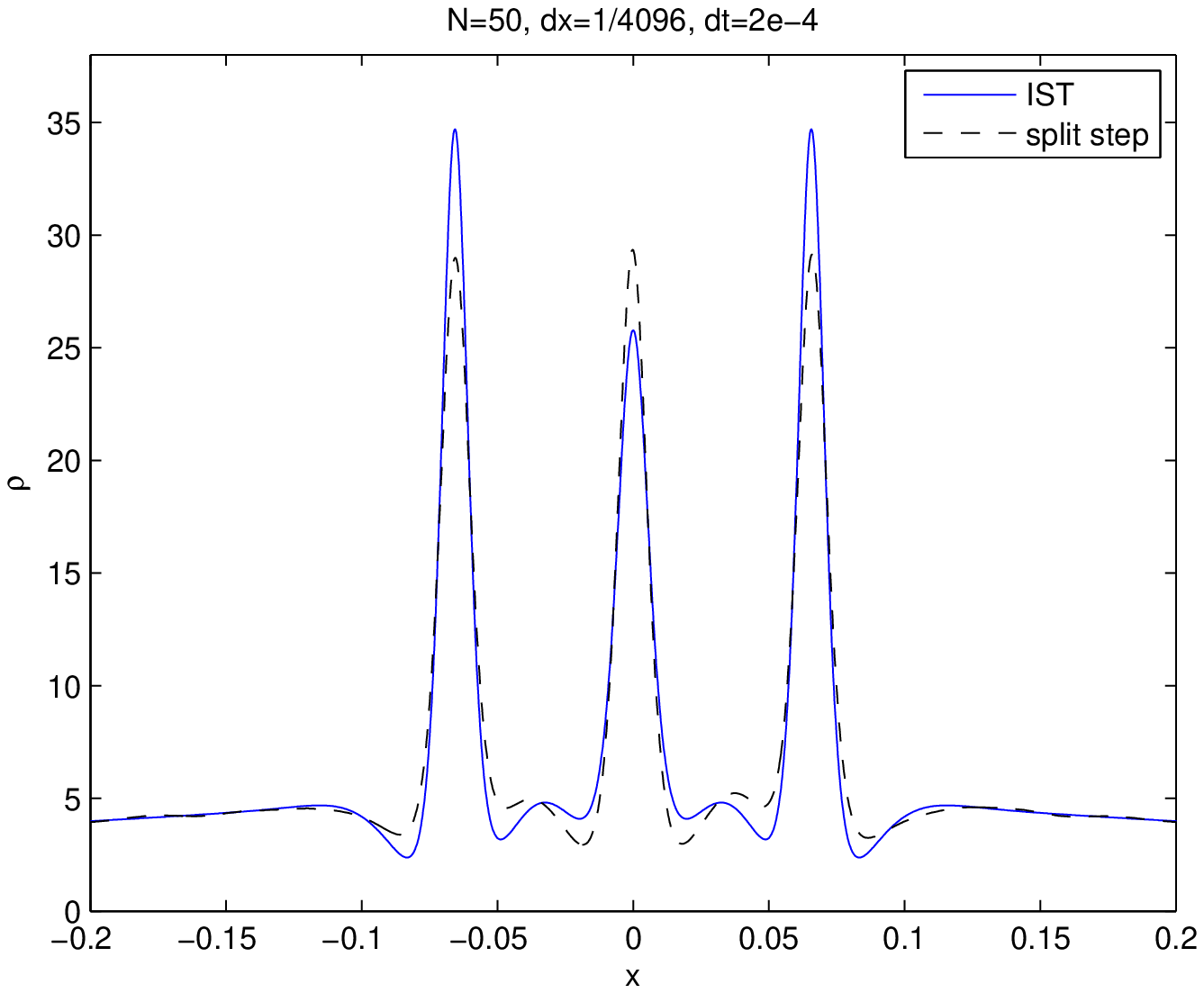}
\includegraphics[width=3.in]{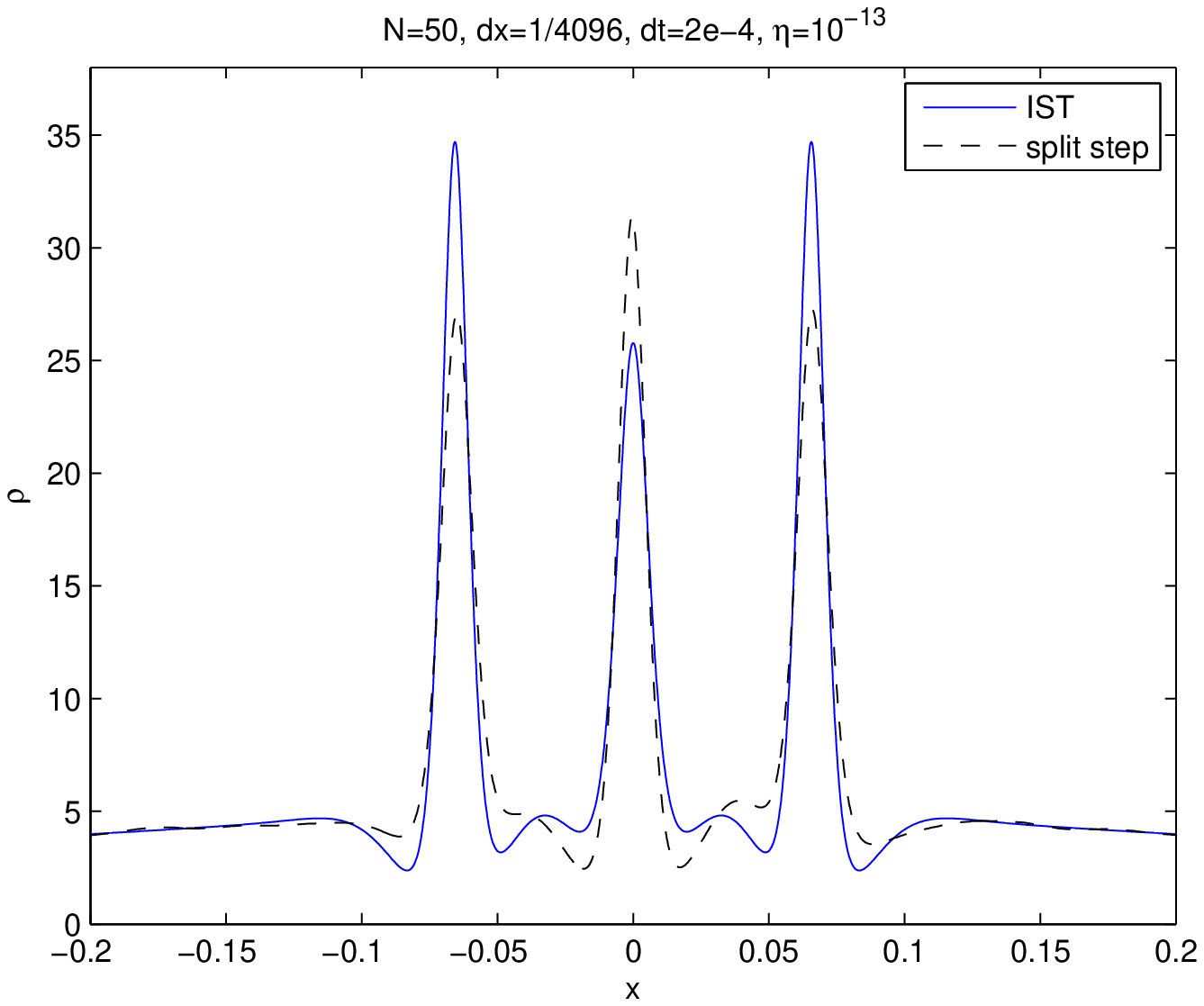}\\
(c)\includegraphics[width=3.in]{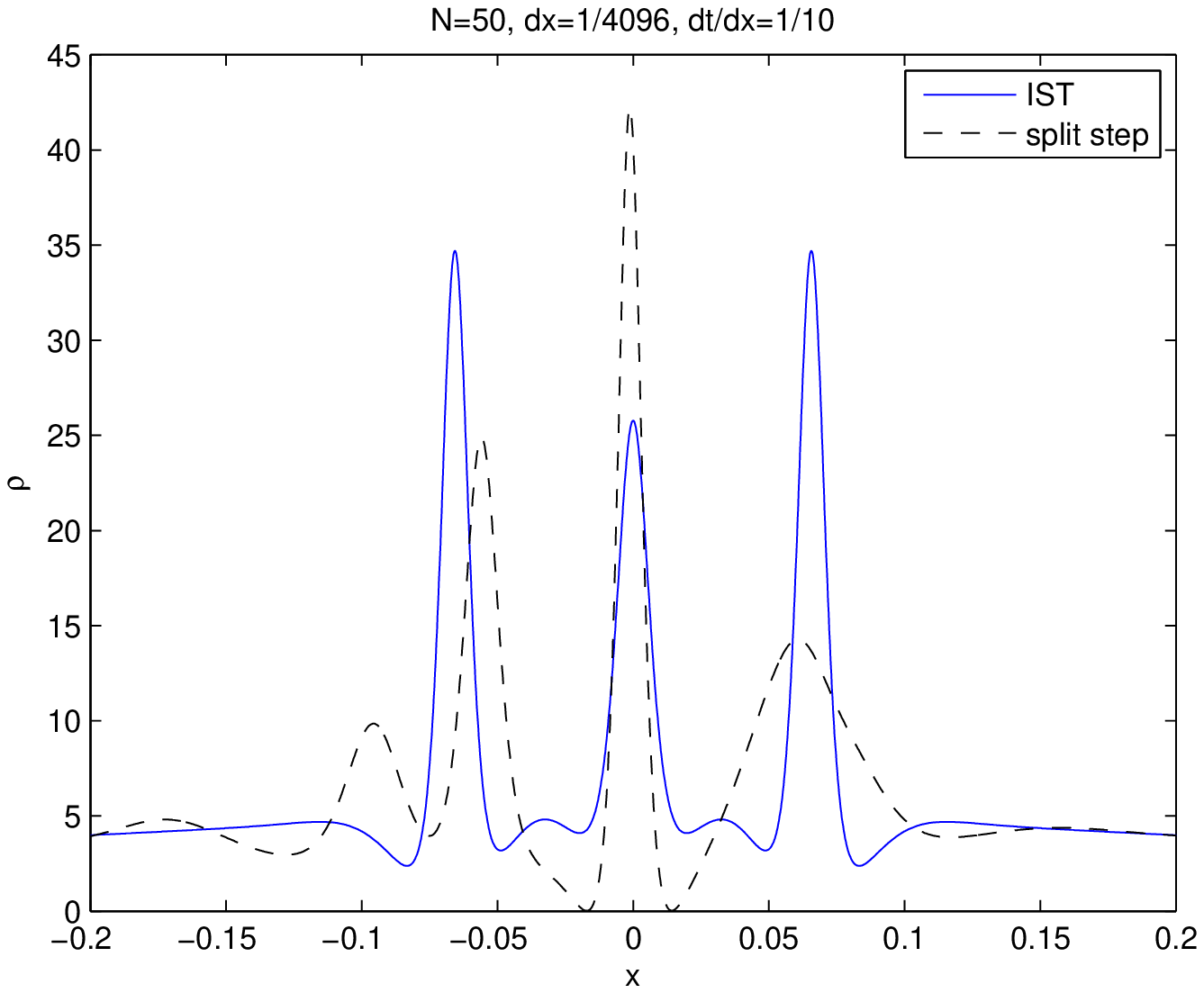}
\includegraphics[width=3.in]{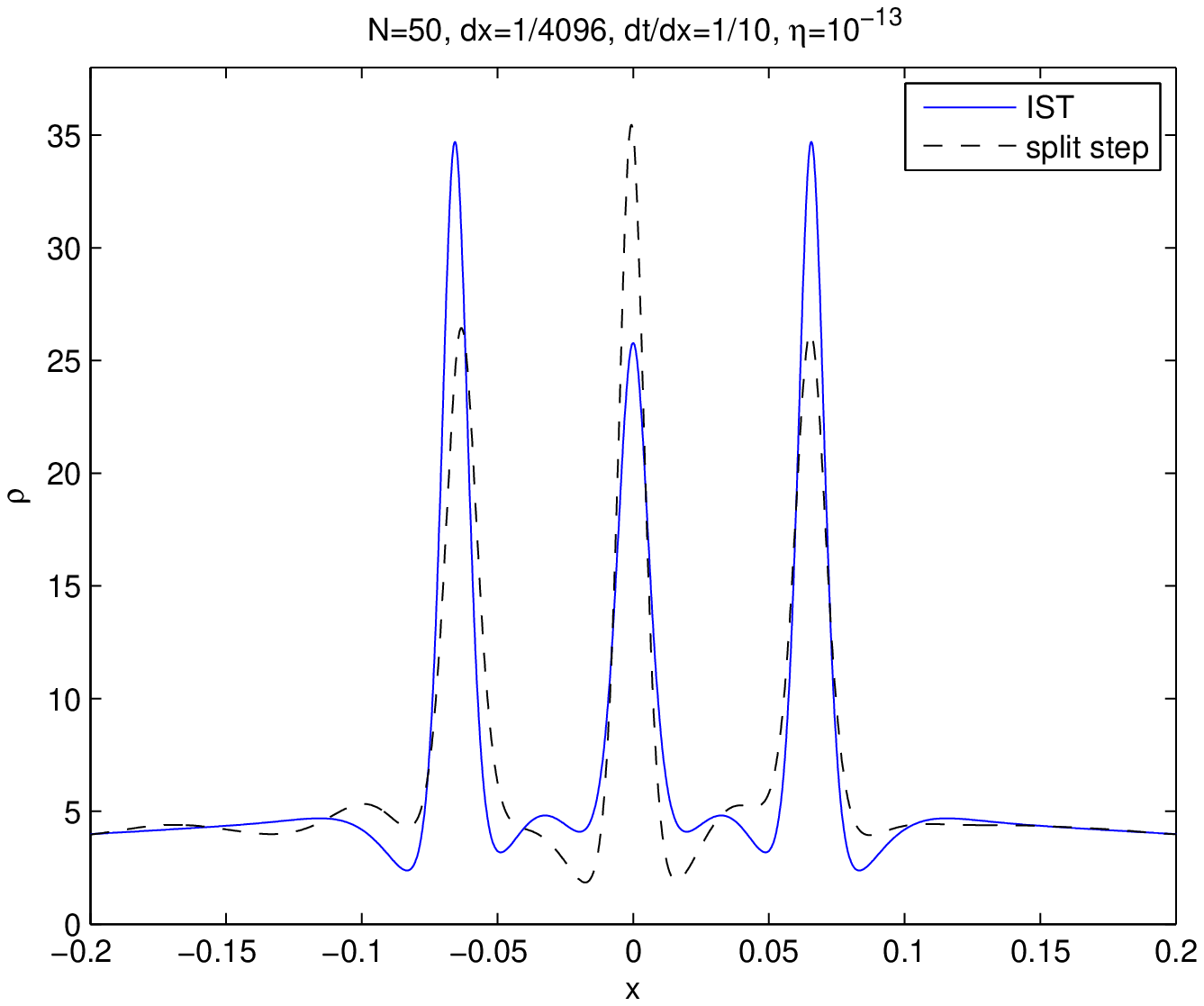}
\caption{$N=50$ at $t=0.3$. Refinement study of the spectral split-step method. Left column: without the Krasny filter. Right column: with the Krasny filter and the threshold is $\eta=10^{-13}$. (a)  $\Delta x=1/2048$, $\Delta t=2\times 10^{-4}$. The $2$-norm error is 0.4397 without the filter and 0.7561 with the filter.  (b)  $\Delta x=1/4096$, $\Delta t=2\times 10^{-4}$. The $2$-norm error is 0.5527 without the filter and 0.7857 with the filter. (c) $\Delta x=1/4096$, $\Delta t=(1/10)\Delta x=2.44140625\times 10^{-5}$. The $2$-norm error is 2.6949 without the filter and 1.177 with the filter.}
\label{fig:N50_SS_t03}
\end{figure}

Similar to Figure \ref{fig:N50_FD_t03}, Figure \ref{fig:N50_SS_t03} is a refinement study of the spectral split-step method. In this study, we explore how temporal and spatial grid sizes used for the spectral split-step method affect the simulations. We also investigate a filtering process introduced by Krasny \cite{Krasny} for ill-posed initial value problems, such as the vortex sheet  roll-up problem. Krasny's experience with vortex sheet simulations suggests that if the problem is ill-posed, fewer grid points should be used for the simulation. The reason for this is that more grid points introduce shorter wavelengths into the numerical solution, and once the short wavelengths are spuriously perturbed by roundoff error, the computation collapses quickly. Krasny proposed a filtering process, now known as the Krasny filter, that eliminated Fourier modes whose amplitudes were smaller than a threshold to restore smoothness of the roll-up. 
For the focusing NLS equation (\ref{eq:nls_ivp}), it is known that  the problem becomes ill-posed when $\epsilon$ approaches to zero \cite{CT}. Hence, the Krasny filter is sometimes applied to simulations of the equation in the semiclassical regime \cite{BJM, CT}.  In particular, Bao et al. \cite{BJM} showed that the Krasny filter successfully restored symmetry for simulations that showed breaking of symmetry. 

In the recent papers of Bao et al. \cite{BJM} and Jin et al. \cite{JMS}, the authors suggested that typically the temporal and spatial grids used for the focusing NLS equation should satisfy 
\beq\label{mesh_size}
\Delta x=O(\epsilon),\quad \Delta t=o(\epsilon)\,.
\eeq
Taking this suggestion and the hint from our numerical experiment for $N=40$ in Figure \ref{fig:N40_t03}(b), we first use the same spatial and temporal grid sizes, $\Delta x=1/4096$ and $\Delta t=(1/10)\Delta x=1/40960$, as that used in Figure \ref{fig:N40_t03}(b) for $N=50$. On the left-hand-side of Figure \ref{fig:N50_SS_t03}(c), we show that using this set of mesh sizes does not produce a satisfactory result, compared with the IST calculation. We apply the Krasny filter to the same calculation so that if $|\hat{u}_k| < \eta$, where $|\hat{u}_k|$ is the amplitude of $\hat{u}_k$ and $\hat{u}_k$ is defined in equation (\ref{eq:FFT}), we manually set $\hat{u}_k$ zero. Here $\eta$ is the threshold level. On the right-hand-side of Figure \ref{fig:N50_SS_t03}(c), we show that when the Krasny filter ($\eta=10^{-13}$) is applied at the end of each time step to the simulation, the filtered result has a better match with the IST calculation. The $2$-norm error is reduced to 1.1177 from 2.6949, and symmetry has mostly been restored. Our numerical experiments show that if we use finer temporal grid sizes, with or without the filtering process, the results only get worse. If we, on the other hand, decrease the temporal grid size, by trial-and-error, we find that when $\Delta t=2\times 10^{-4}$, without the filtering process, we obtain a reasonable match with the IST result, as shown on the left-hand-side of Figure \ref{fig:N50_SS_t03}(b). If we keep this temporal grid size and coarsen the spatial grid size to $\Delta x=1/2048$, without the filtering process, we obtain an even better match with the IST calculation, as shown on the left-hand-side of Figure \ref{fig:N50_SS_t03}(a). On the right-hand side of Figure \ref{fig:N50_SS_t03}(a) and (b), we show that the Krasny filter does not improve the results. On the contrary, more significant phase-shift is observed for calculations with the Krasny filter. We note that we obtain almost identical results for the threshold levels from $10^{-15}$ to $10^{-10}$. In the next numerical experiment, we further show that the Krasny filters not only do not improve the results, they produce results that do not match with the IST calculations for small $\epsilon$.\\

\noindent
{\bf Example 3:} 
$
u(x,0) = 2\sech(x),\,\, N=54,\,\, \epsilon=1/27\approx 0.037,\,\,\text{final time}\,\, t=0.3.\\
$

\begin{figure}[hbtp]
\centering
(a)\includegraphics[width=3in]{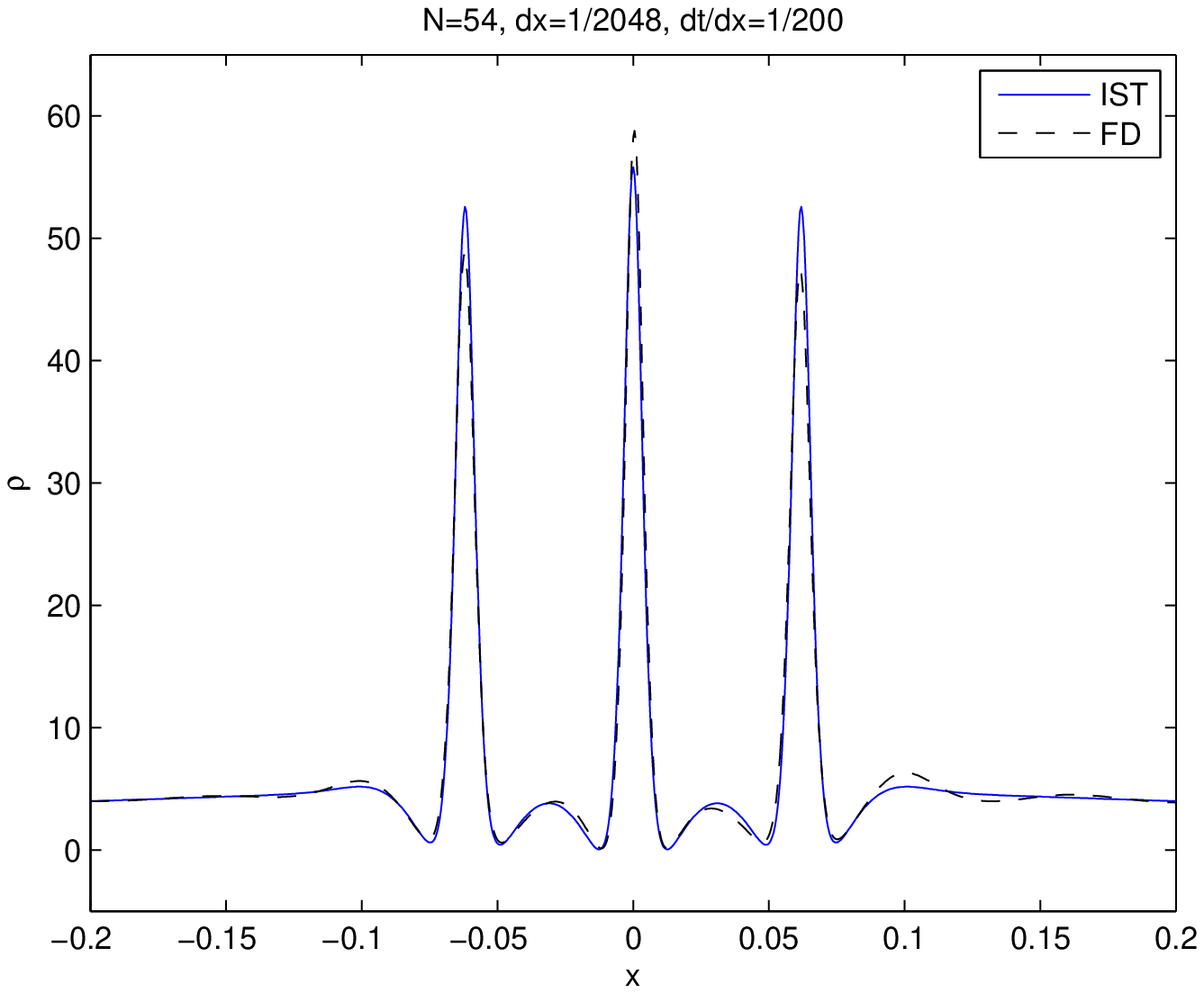}
(b)\includegraphics[width=3in]{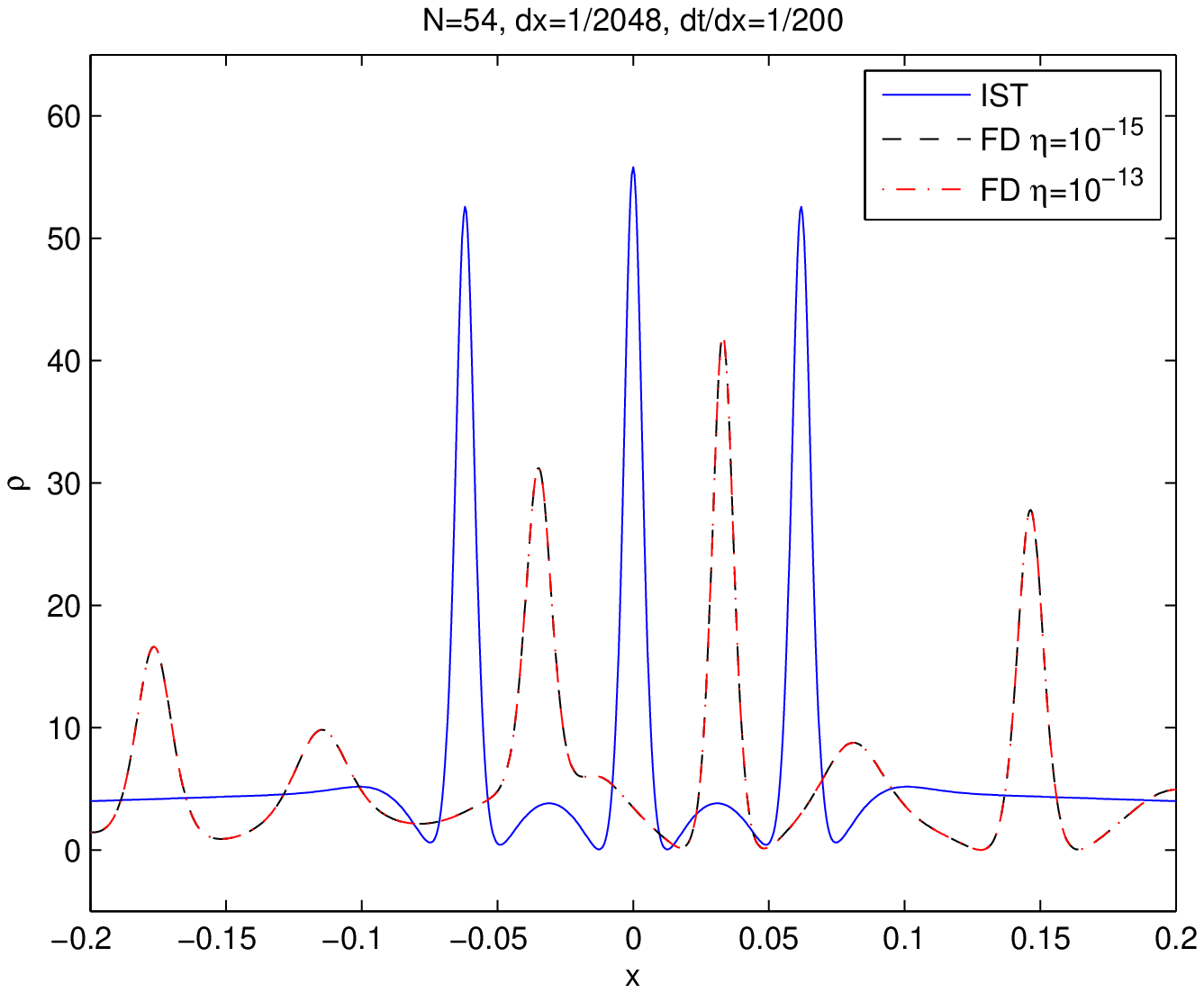}
\caption{$N=54$, $\Delta x=1/2048$, $\Delta t / \Delta x =1/200$, $t=0.3$. Computation by using the finite difference method. (a) Without the Krasny filter. (b) With the Krasny filter. The thresholds are $\eta=10^{-15}$ and $10^{-13}$, respectively.}
\label{fig:N54_FD_t03}
\end{figure}

In this example, we show that for $N=54$ ($\epsilon\approx 0.037$), the proposed finite difference method captures the proper waveform of the solution, while the solution obtained by using the spectral split-step method is heavily influenced by numerical artifacts. We further show that the Krasny filter not only fails to reduce the numerical artifacts in both methods, but produces solutions that are drastically different from the IST calculation. Figure \ref{fig:N54_FD_t03}(a) shows that without the Krasny filter, the finite difference method produces a solution that match the IST result closely. We note that the spatial grid size is $\Delta x=1/2028$ in this simulation. Using either a finer grid such as $\Delta x=1/4096$ or a coarser grid such as $\Delta x=1/1024$ will degenerate the result, regardless the choice of temporal grid size, based on our numerical experiments. The convergence tolerance is $\gamma=10^{-15}$ for the simulation. Figure \ref{fig:N54_FD_t03}(b) shows the same computation, except Krasny's filter ($\eta=10^{-13}$ and $10^{-15}$) is applied at the end of each time step. The numerical results are identical for these two different thresholds, and they fail to match the IST calculation.  

\begin{figure}[hbtp]
\centering
(a)\includegraphics[width=3in]{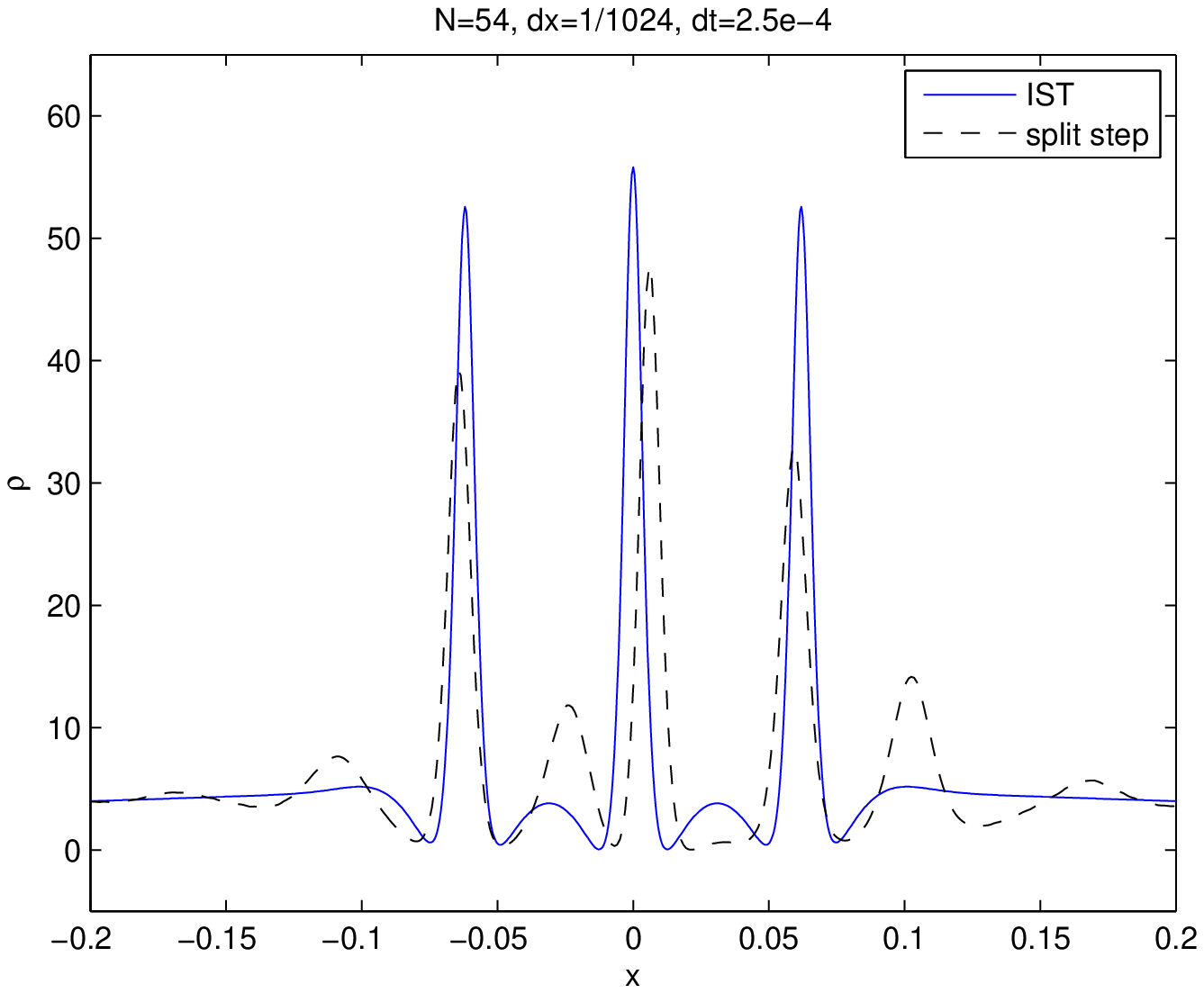}
(b)\includegraphics[width=3in]{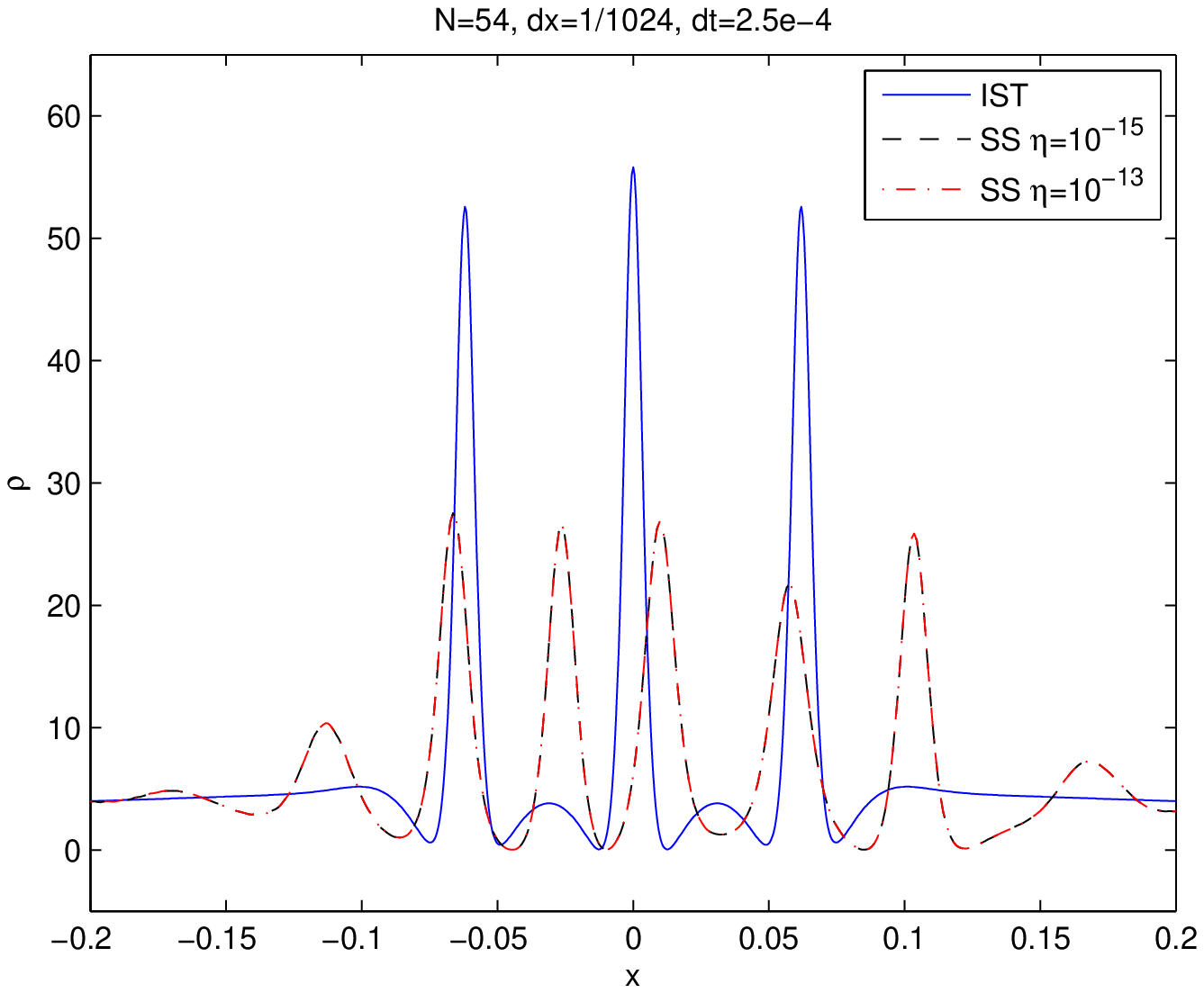}
\caption{$N=54$, $\Delta x=1/1024$, $\Delta t =1200$, $t=0.3$. Computation by using the spectral split-step  method. (a) Without the Krasny filter. (b) With the Krasny filter. The thresholds are $\eta=10^{-15}$ and $10^{-13}$, respectively.}
\label{fig:N54_SS_t03}
\end{figure}

Figure \ref{fig:N54_SS_t03}(a) shows that the result by using the spectral split-step method does not match the IST calculation for $N=54$. The spatial grid size is $\Delta x=1/1024$ in this simulation. Similar to the finite difference method, using  either a finer grid such as $\Delta x=1/2048$ or a coarser grid such as $\Delta x=1/512$ will not improve the result.  When the Krasny filter is applied to this simulation, the result becomes even worse, as shown in Figure \ref{fig:N54_SS_t03}(b). Similar to the finite difference method, two thresholds, $\eta=10^{-13}$ and $10^{-15}$, are applied to the computation, and the outcomes are identical. 

The results of this numerical experiment are hardly surprising.  As pointed out by Jin et al. \cite{JMS}, the Krasny filtering process actually violates the conservation of mass, which means that the cut small-amplitude Fourier modes could very much be part of the solution. 

\section{Evolution: $A_0$ versus $u_0^{(\eps)}$}\label{sec:gauss}

\begin{figure}[hbtp]
\centering
(a)\includegraphics[width=3in]{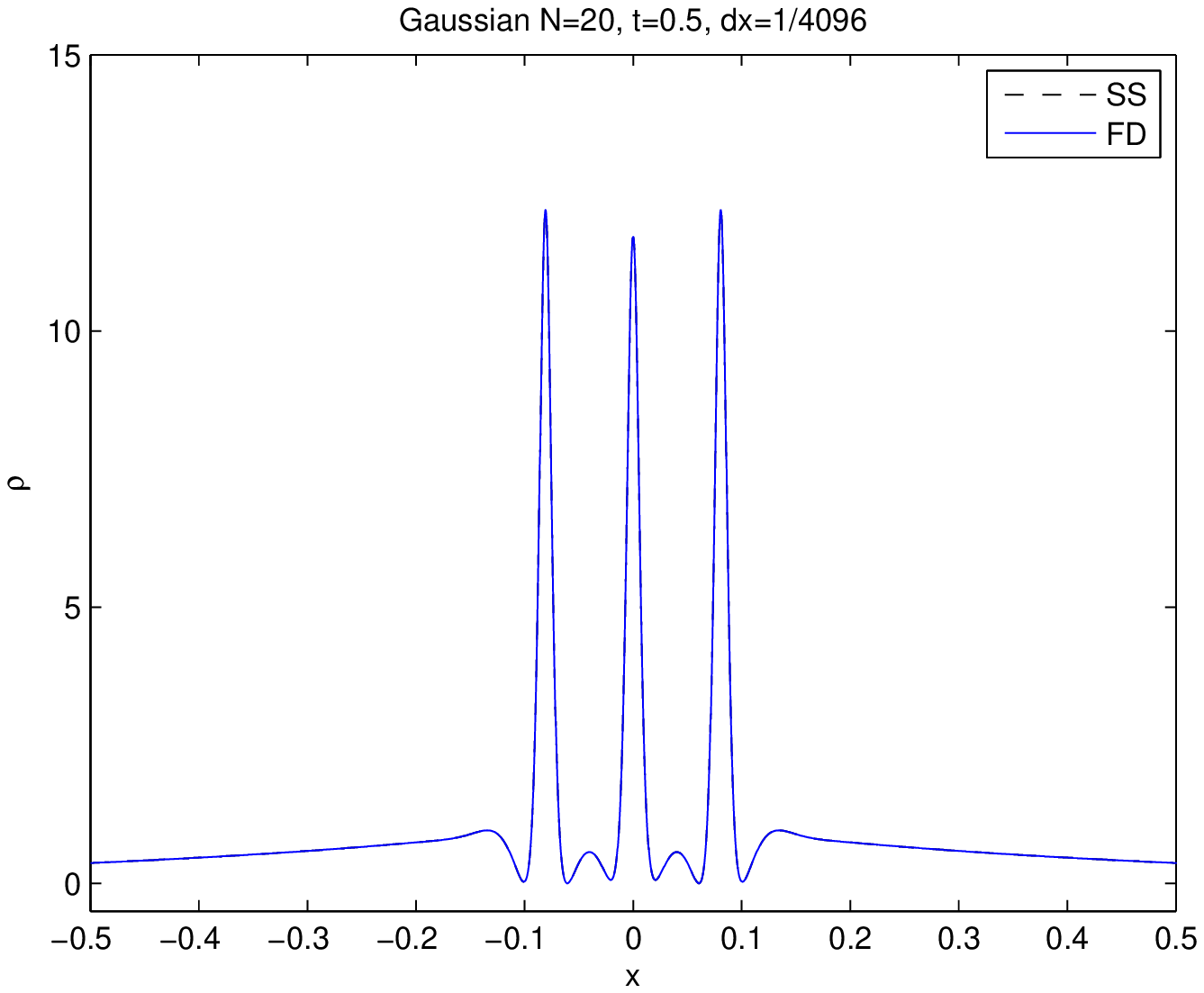}
(b)\includegraphics[width=3in]{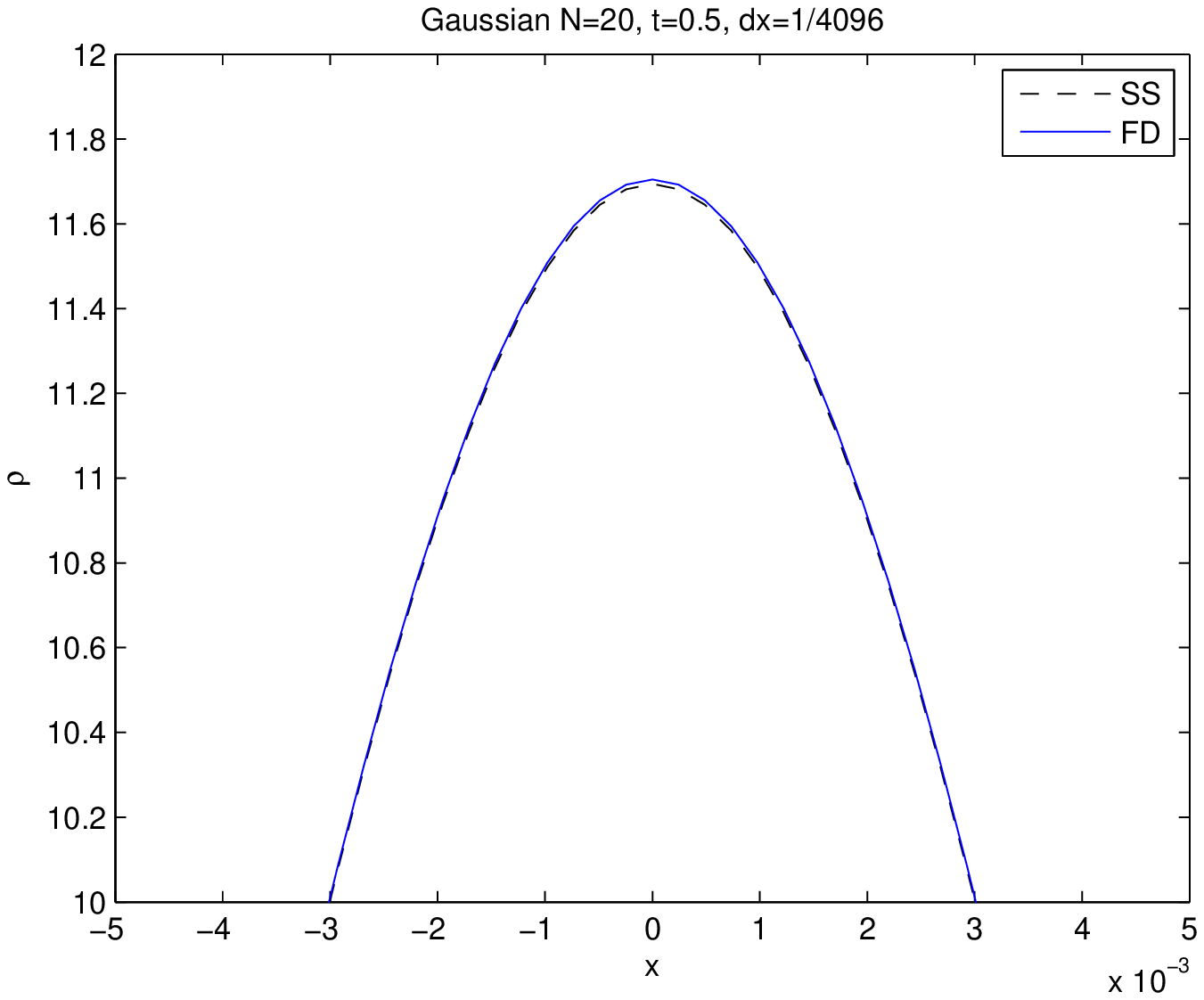}
\caption{Gaussian data for $N=20\, (\epsilon\approx 0.0282)$ at final time $t=0.5$. The grid resolution is $\Delta x =1/4096$. (a) Comparison between the finite difference method and the split-step method. (b) Magnification of the centered peak in (a).}
\label{fig:N20_gaussian}
\end{figure}

In this section, we compare numerical solutions obtained by the evolutionary numerical methods discussed in Section \ref{sec:methods} and by the IST for the Gaussian SSE described in Section \ref{sec:wkb}. First, for the initial data given in equation (\ref{eq:gauss}), we compare the numerical solutions obtained by the finite difference method and the split-step method to benchmark the solutions for this initial data. Table \ref{tab:gaussian_FD_SS} shows the 2-norm difference of $\rho$ between these two methods at final times $t=0.1,\,0.2,\,0.3,\,0.4,\,$ and $0.5$, with number of solitons, $N=5,\,10,\,15$, and $20$. We recall that the relation between the small parameter $\epsilon$ and the number of solitons $N$ for this Gaussian initial data is given in equation (\ref{eq:epsilon_N}). The table shows that at this range of $\epsilon$ and final times, numerical solutions obtained by these two methods follow each other closely. The computational domain is $-10\le x\le 10$ and periodic. The grid resolution is $\Delta x =1/4096$ for both methods, and the difference is measured for $-1\le x\le 1$. Figure \ref{fig:N20_gaussian} shows the case that has the largest difference in Table \ref{tab:gaussian_FD_SS}, for which $N=20$ and the final time $t=0.5$. The two results are visually indistinguishable under the scale as shown in (a), and there is a moderate visual discrepancy after we magnify the graph, as shown in (b), for the centered peak.  We remark that our numerical investigation for the Gaussian initial data below is limited to these ranges of $N$ and $t$. 

\begin{table}[htpbt]
\caption{2-norm difference of $\rho$ between the finite difference and the split-step methods for the Gaussian initial data.}
\label{tab:gaussian_FD_SS}
\centering
\begin{tabular}{c|ccccc}
\hline
\backslashbox{$N$}{$t$} & 0.1 &  0.2 & 0.3 & 0.4 & 0.5 \\\hline
5 &3.8360E-8 &1.8054E-7 &5.8820E-7 &2.3012E-6 &1.9354E-5  \\\hline 
10&1.6094E-7 &7.7978E-7 &2.7336E-6 &2.0142E-5 &1.1714E-4  \\\hline
15 &3.6547E-7 & 1.7772E-6 & 6.3417E-6 &7.5820E-5 &5.6675E-4\\\hline
20 &6.5206E-7& 3.1748E-6 & 1.1401E-5 & 2.1776E-4 &1.7275E-3 \\\hline
\end{tabular}
\end{table}

\begin{figure}[hbtp]
\centering
(a)\includegraphics[width=3in]{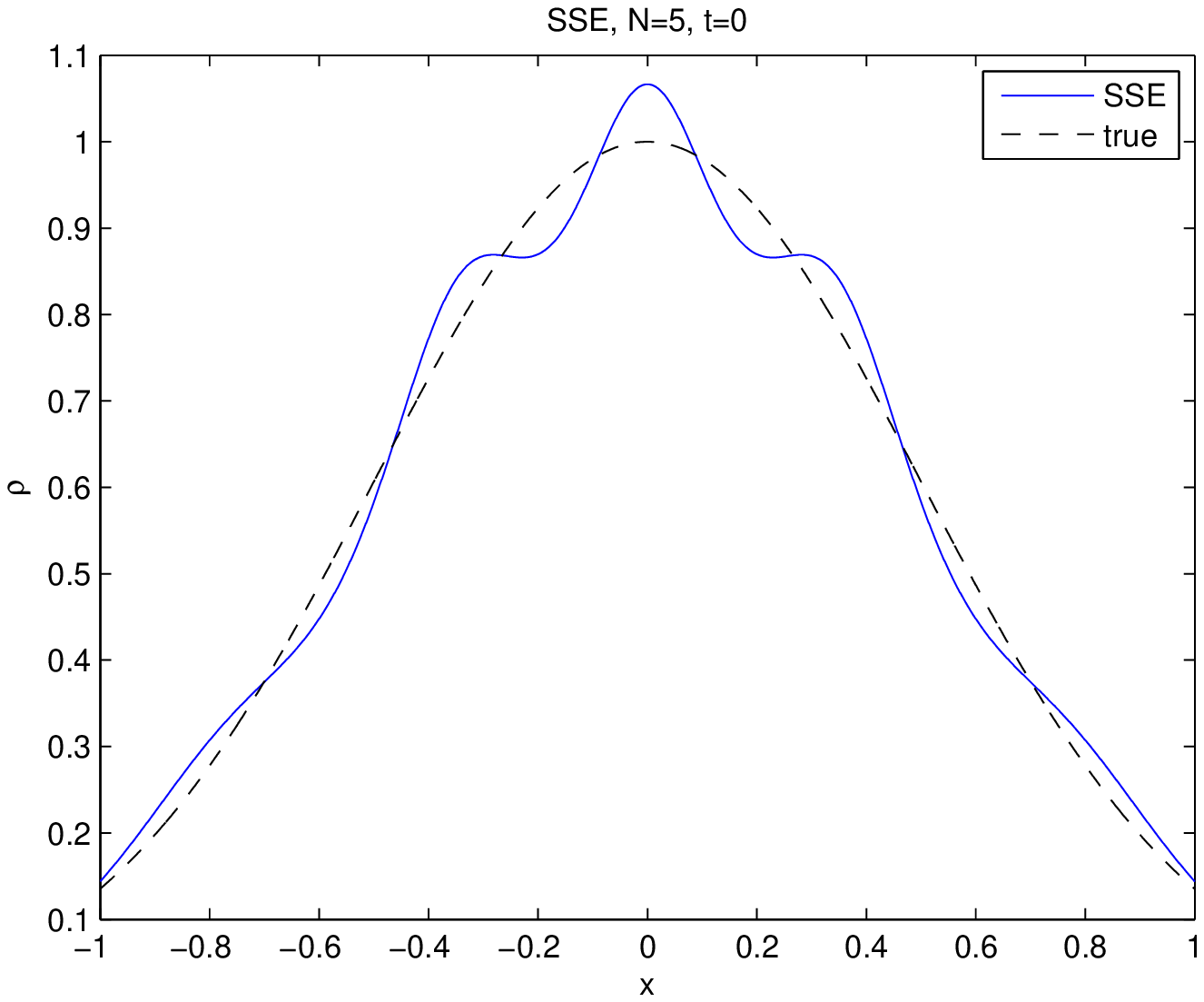}
(b)\includegraphics[width=3in]{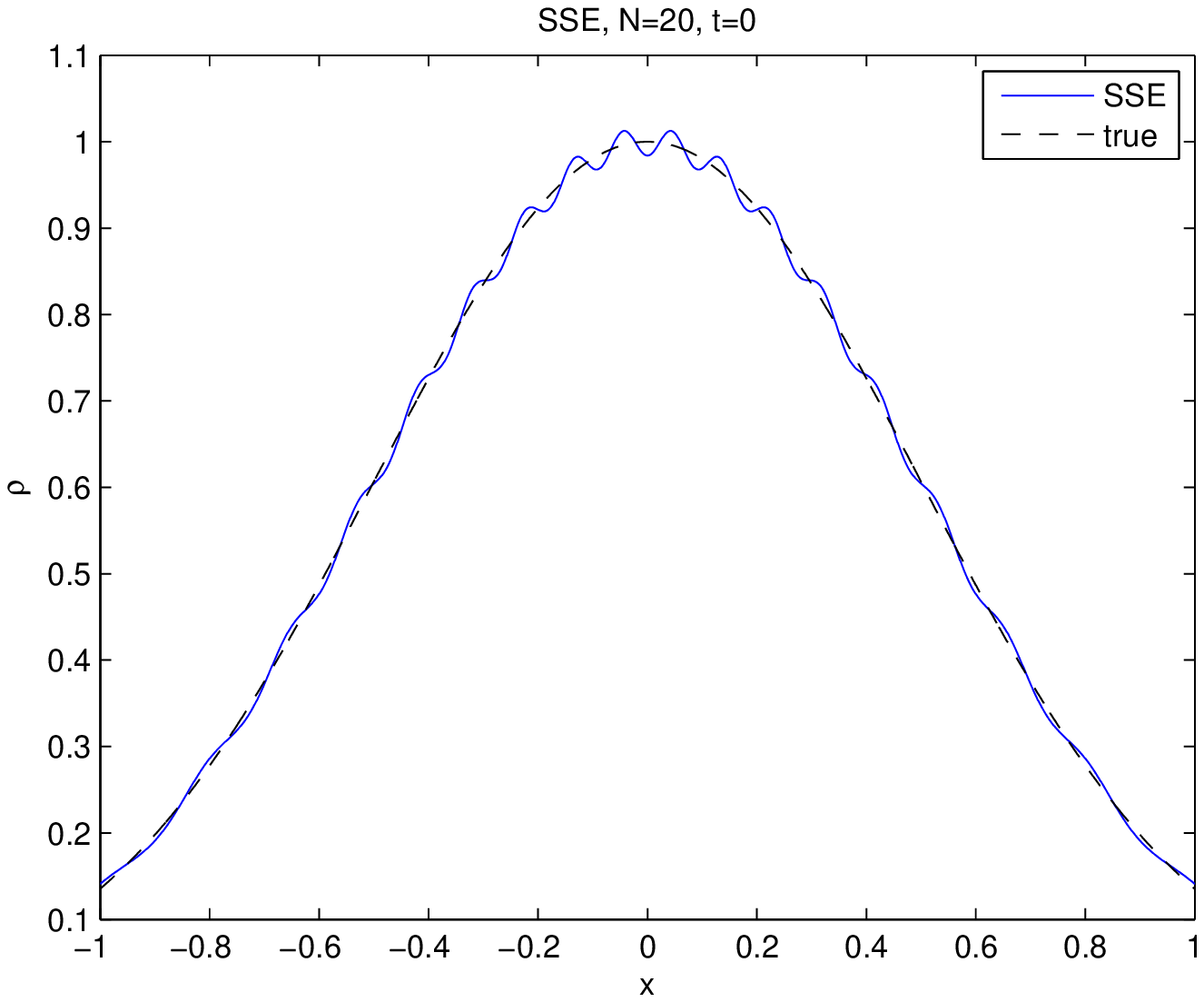}
\caption{Reconstruction of the Gaussian initial data by the Gaussian SSE. (a) $N=5$, (b) $N=20$.}
\label{fig:gaussian_initial}
\end{figure}

\begin{figure}[hbtp]
\centering
(a)\includegraphics[width=3in]{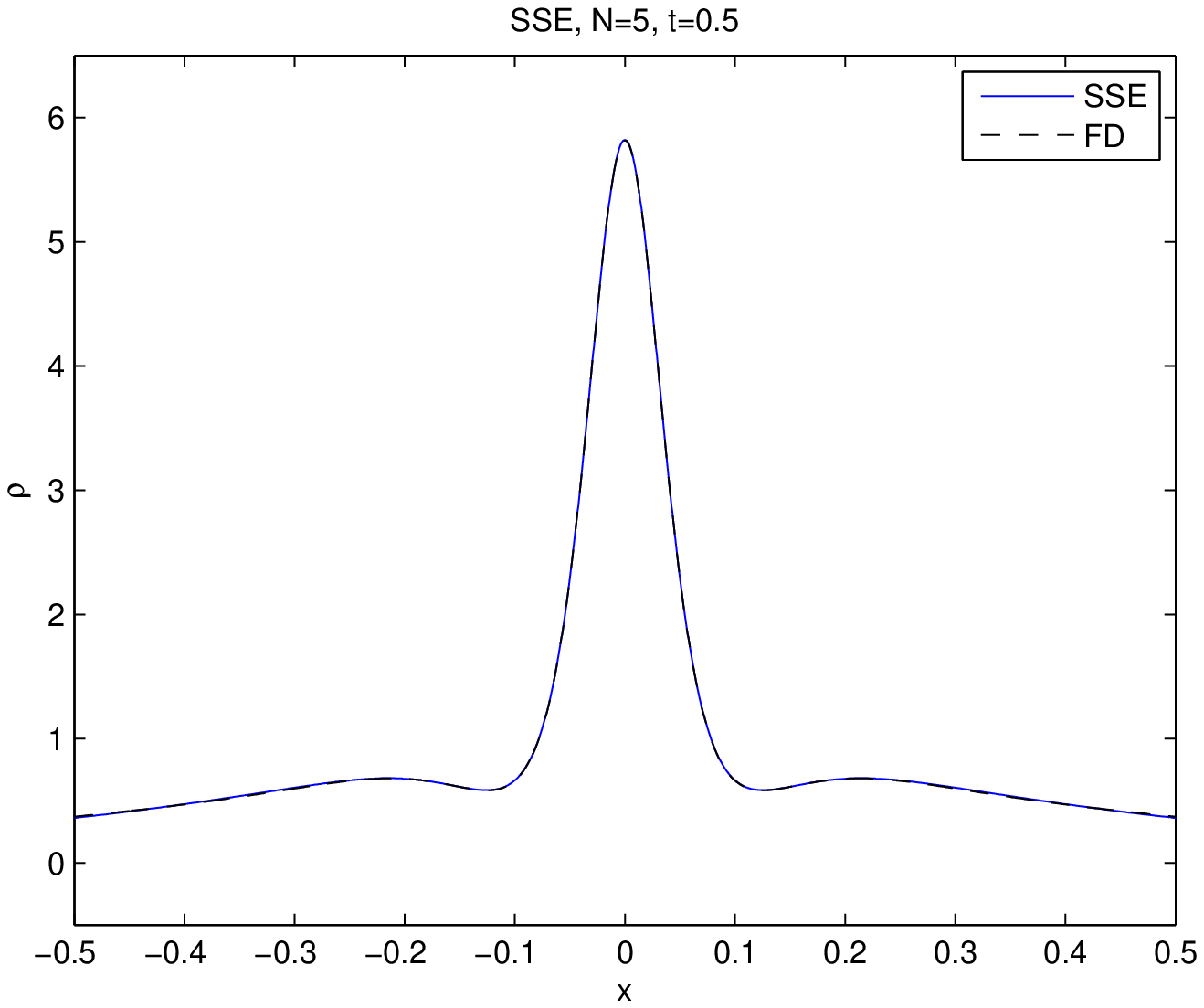}
(b)\includegraphics[width=3in]{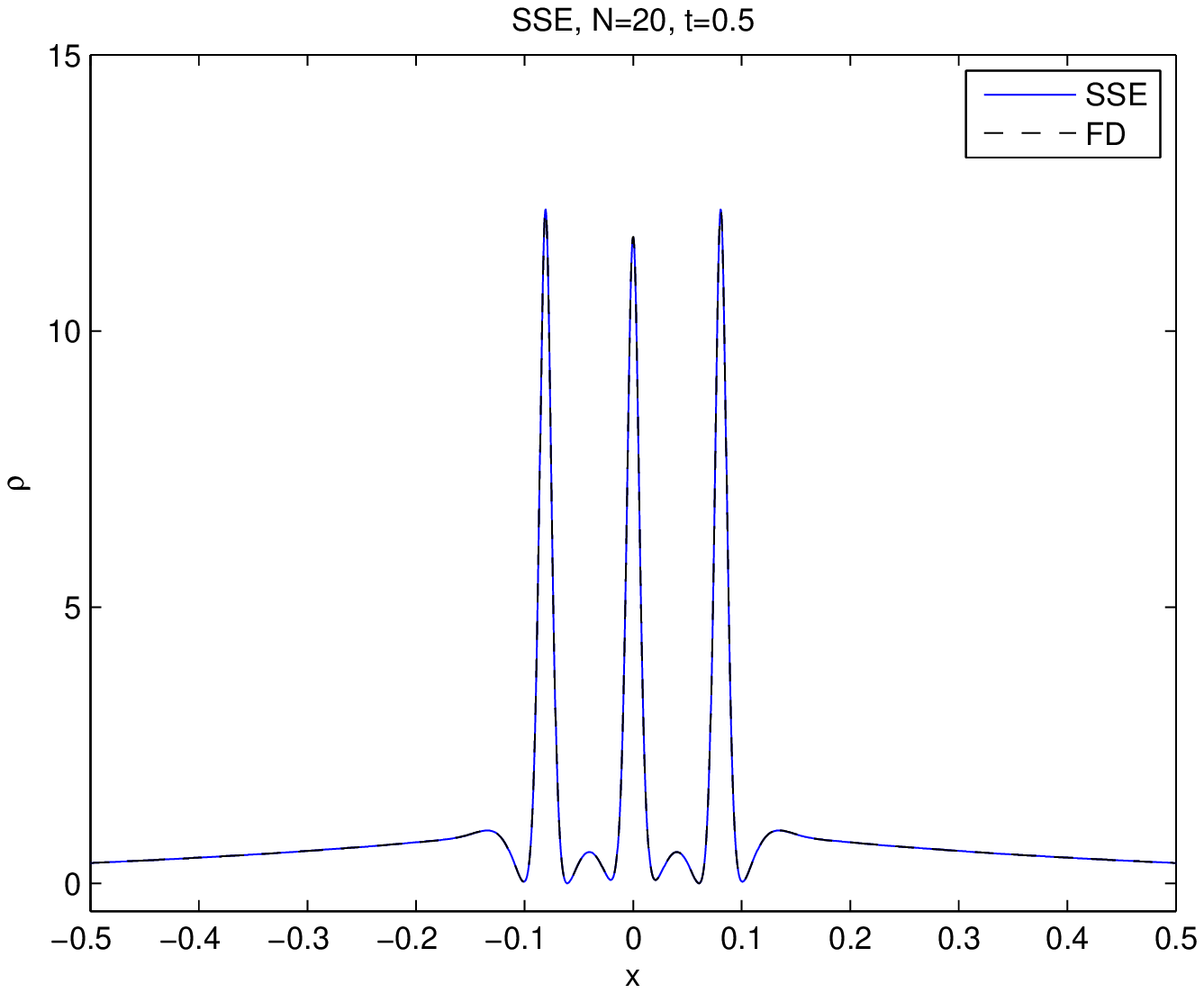}
\caption{Comparison of a member of the Gaussian SSE (computed via IST) with the corresponding finite-difference solution with Gaussian initial data at $t=0.5$. (a) $N=5$, (b) $N=20$.}
\label{fig:gaussian_t05}
\end{figure}

Figure \ref{fig:gaussian_initial} shows the reconstruction of the initial data associated with the Gaussian SSE, and its comparison with the true Gaussian initial data for $N=5$ and $20$, respectively. These reconstructed initial data can be seen as perturbation of the true Gaussian data. Our main interest is to understand how these perturbed data evolve with time in comparison to the evolution of the true data. 
For large $N$ (small $\epsilon$), while the initial perturbation is small, how the perturbed data will evolve with time in comparison with the true evolution is not clear. In principle, the evolution of the perturbed data could depart from the true evolution in a short time, if $\epsilon$ is small enough.  Figure \ref{fig:gaussian_t05} shows that at time $t=0.5$, the perturbed data become visually indistinguishable from the finite-differenced solution for both $N=5$ and $N=20$. The computational domain is $-10\le x\le 10$ and periodic, with the grid resolution $\Delta x =1/4096$ for the finite difference method.

\begin{figure}[hbtp]
\centering
\includegraphics[width=4in]{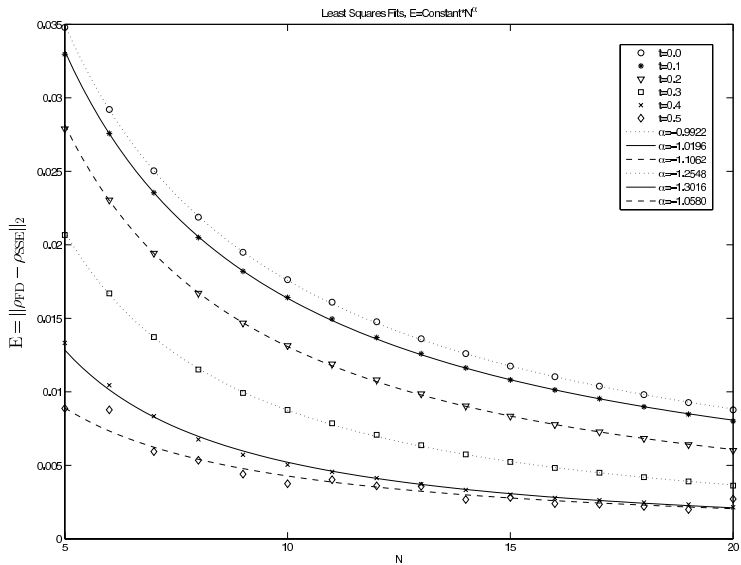}
\caption{The 2-norm differences versus $N$ for $t=$ 0.0, 0.1, 0.2, 0.3, 0.4, and 0.5. The data points are the computed values of the 2-norm error $\mathrm{E}=\|\rho_\mathrm{FD}-\rho_\mathrm{SSE}\|_2$ from Table \ref{tab:gaussian_FD_IST}. The plotted curves are of the form $\mathrm{E}=C\cdot N^\alpha$ where the constants $C$, $\alpha$ are determined by a least squares fit to the data points; see Table~\ref{tab:least_data}. The values for $\alpha$ in the legend show a $O(1/N)=O(\eps)$ rate of convergence even for the times $t=0.4$ and $t=0.5$ which are after the breaking time.}
\label{fig:gaussian_decay_N}
\end{figure}

We now systematically compute the 2-norm difference of $\rho$ between the SSE solutions and those of the finite difference method (representing the true evolution of the initial data $u_0=\exp(-x^2)$) for the ranges of $N=5$ to $N=20$ and $t=0.0$ to $t=0.5$.  Table \ref{tab:gaussian_FD_IST} shows that the differences between the two solutions diminish with time for all $N$ (except the last point, when $N=20$, $t=0.5$). Figure \ref{fig:gaussian_decay_N} represents the same data; the markers show the 2-norm differences versus $N$ for times $t=$ 0.0, 0.1, 0.2, 0.3, 0.4, and 0.5, respectively. The figure shows a remarkable consistency in the decay of the error as $N$ increases; we discuss this further in the next section. 
\begin{table}[htpbt]
\caption{2-norm difference of $\rho$ between the finite-difference solution (for $t>0$) with Gaussian initial data and the corresponding member of the Gaussian SSE. }
\label{tab:gaussian_FD_IST}
\centering
\begin{tabular}{c|cccccc}
\hline
\backslashbox{$N$}{$t$} & 0 & 0.1 &  0.2 & 0.3 & 0.4 & 0.5 \\\hline
5 &3.4786E-2 &3.2965E-2 &2.7883E-2 &2.0672E-2 &1.3330E-2 &8.8712E-3 \\\hline 
6 &2.9209E-2 &2.7575E-2 &2.3040E-2 &1.6697E-2 &1.0454E-2 &8.7774E-3  \\\hline 
7 &2.5037E-2 &2.3543E-2 &1.9420E-2 &1.3727E-2 &8.3449E-3 &5.9334E-3  \\\hline 
8 &2.1881E-2 &2.0498E-2 &1.6701E-2 &1.1517E-2 &6.7653E-3 &5.3467E-3  \\\hline 
9 &1.9490E-2 &1.8200E-2 &1.4677E-2 &9.9244E-3 &5.7163E-3 &4.4058E-3 \\\hline 
10 &1.7628E-2 &1.6419E-2 &1.3136E-2 &8.7682E-3 &5.0452E-3 &3.7535E-3  \\\hline 
11 &1.6095E-2 &1.4959E-2 &1.1889E-2 &7.8615E-3 &4.5615E-3  &4.0257E-3 \\\hline
12 &1.4772E-2 &1.3698E-2 &1.0812E-2 &7.0770E-3 &4.1389E-3  &3.6043E-3 \\\hline
13 &1.3611E-2 &1.2591E-2 &9.8614E-3 &6.3696E-3 &3.7160E-3  &3.5731E-3  \\\hline
14 &1.2605E-2 &1.1631E-2 &9.0362E-3 &5.7488E-3 &3.3310E-3  &2.6767E-3  \\\hline
15 &1.1751E-2 &1.0818E-2 &8.3428E-3 &5.2359E-3 &3.0135E-3  &2.8144E-3\\\hline
16 &1.1025E-2 &1.0131E-2 &7.7671E-3 &4.8292E-3 &2.7846E-3  &2.4003E-3  \\\hline
17 &1.0389E-2 &9.5319E-3 &7.2736E-3 &4.4972E-3 &2.6298E-3  &2.3538E-3 \\\hline
18 &9.8091E-3 &8.9850E-3 &6.8248E-3 &4.1976E-3 &2.4770E-3  &2.2044E-3 \\\hline
19 &9.2696E-3 &8.4754E-3 &6.4019E-3 &3.9052E-3 &2.3439E-3  &1.9980E-3 \\\hline
20 &8.7755E-3 &8.0077E-3 &6.0095E-3 &3.6229E-3 &2.1483E-3  &2.7102E-3 \\\hline
\end{tabular}
\end{table}
%
%
We also note that the 2-norm difference between the finite difference and the split-step method is of the order $10^{-3}$ for $N=20$ and $t=0.5$, and this is also the difference between the finite difference method and the IST. Therefore, at this point it is difficult to determine whether the difference between the finite difference method and the IST truly represents the difference between the true solution and the IST-generated member of the SSE. Also, it has little meaning to do more comparison for larger $t$ or $N$ beyond this point. 


\section{Discussion and Concluding Remarks}\label{sec:discuss}

We have compared a spectral split-step method and an implicit finite difference method for solving the focusing NLS equation in the semiclassical regime. In the special case that the initial data is $A_0(x)=A\sech(x)$, the IST solution serves as an exact solution for the comparison. We find that the spectral split-step method is more efficient compared with the proposed implicit finite difference method. However, for small $\epsilon$, we find that the proposed implicit finite difference method is less sensitive to the choice of spatial and temporal grid sizes than the split-step method; poor choices lead to numerical artifacts caused by numerical roundoff error. We observe that to obtain simulations with the fewest numerical artifacts for the $N$-soliton problem with large $N$ (e.g.  $N \ge 54$ and $A=2$), the use of spatial and temporal grid sizes for the numerical methods (both split-step method and the proposed finite difference method) should follow Krasny's suggestions \cite{Krasny}, i.e., use fewer grid points and larger time-step sizes, rather than the meshing strategy in equation (\ref{mesh_size}).  

We also investigated a filtering process, known as the Krasny filter. We find that the process may help to restore symmetry of the solution, but the restored solution may not represent the solution of the problem. Furthermore, when $\epsilon$ is small, the filtering process could even destroy good numerical simulations of the problem. This is because for small $\epsilon$, the highly oscillatory analytical solution could be a superposition of those small-amplitude Fourier modes, it is not possible for the filter to distinguish the small-amplitude Fourier modes of the solution from those due to roundoff error.   

\begin{figure}[hbtp]
\centering
\includegraphics[width=3in]{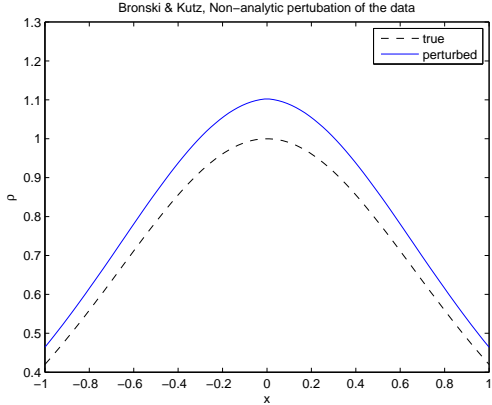}
\caption{The non-analytic perturbation of Bronski \& Kutz \cite{BK}.}
\label{fig:bk_data}
\end{figure}

Finally, we used the two studied numerical methods to investigate the Gaussian SSE for the focusing NLS equation. Within the range of $\epsilon$ and $t$ for which we are confident about our numerical solutions, we find that for larger $\epsilon$, the perturbation of the initial data is quickly dissipated and the SSE solution becomes close to the finite differenced solution, a good approximation to the true solution. 
We see this as a reflection of the particularly special nature of the perturbations we consider here; after all, they are connected to the data through the WKB analysis of \eqref{eq:zs}. By way of comparison, we recall one of the experiments of Bronski \& Kutz \cite{BK} which featured a non-analytic perturbation of initial data. In particular, they considered a perturbation, see Figure~\ref{fig:bk_data}, of the initial data 
$
u_0(x)=\sech(x)
$
by a small multiple of the tent function  
\beq
f(x)=\begin{cases}
1-\left|\frac{1}{3}x\right|\,, &\text{if}\; |x|<3, \\
0\,, & \text{if}\;|x|\geq 3\,.
\end{cases}
\eeq
In this case, Bronski \& Kutz found that the evolution in their simulations was extremely sensitive to even small amounts of non-analyticity. Similarly, Clarke \& Miller \cite{CM} also found that the problem is quite sensitive to non-analytic perturbations; they found wild behavior even for small perturbations of class $C^2(\RR)$.

To reveal further structure in the data assembled in Table ~\ref{tab:gaussian_FD_IST}, we postulate a relationship for the 2-norm error $\mathrm{E}$ as function of $N$ of the form
\[
\mathrm{E}=\|\rho_\mathrm{FD}-\rho_\mathrm{SSE}\|_2=C\cdot N^\alpha\,,
\]
and we use the data compute values for $C$ and $\alpha$ by least squares. The resulting values are shown in Table \ref{tab:least_data}, and the corresponding curves are plotted in Figure~\ref{fig:gaussian_decay_N}.
\begin{table}
\centering
\caption{The values of $C$ and $\alpha$ for each $t$.}
\label{tab:least_data}
\begin{tabular}{c|ccccccc}
$t$ & 0.0 & 0.1 & 0.2 & 0.3 & 0.4 & 0.5 \\ \hline
$\alpha$ & $-0.9922$ & $-1.0196$ & $-1.1063$ & $-1.2548$ & $-1.3016$ & $-1.0580$ \\
$C$ & $0.1727$ & $0.1713$ & $0.1671$ & $0.1574$ & $0.1044$ & $0.0489$ 
\end{tabular}
\end{table}
The values for $\alpha$ in Table~\ref{tab:least_data} show that the modified initial data arising from the WKB analysis of \eqref{eq:zs} converges to the true data at a $O(1/N)$ rate. (Recalling, \eqref{eq:epsilon_N}, we see equivalently, that this rate of convergence is $O(\eps)$.) Remarkably, our experiments show this rate of convergence is preserved at later times suggesting the possibility of a limited kind of well-posedness in the semiclassical limit. Also, we recall that Clarke \& Miller \cite{CM} showed how to extract an upper bound for the breaking time---roughly, the first time that $|u(\cdot,t)|^2$ begins to exhibit $\eps$-scale oscillations---by an analysis of the formula
\beq\label{eq:break}
|t|=\frac{2}{\pi\sqrt{\rho_\circ}}\int_0^{-\mi A_0^{-1}(\sqrt{\rho_\circ})} E\left(1-\frac{A_0(\mi y)^2}{\rho_\circ}\right)\,\dif y\,,
\eeq
where $E(m)$ is a complete elliptic integral of the second kind, and $A_0$ is the initial data. Using their technique, we compute the first critical point of $t(\rho_0)$ and find an upper bound on the breaking time  for $A_0=\exp(-x^2)$ to be
\[
t_\mathrm{ub}=0.377417\,.
\] 
Thus, our experiments show that the $O(1/N)$ rate of convergence that we see for small times persists even past wave breaking. This we see as especially noteworthy. Taken together, these results are consistent with the conclusion that, despite modulational instability, the asymptotically small modification of the initial data used by Kamvissis et al. \cite{KMM} does not affect the semiclassical limit.   
Still, it is difficult, based on our experiments, to draw definitive conclusions about the true limiting behavior. For our approach to yield more definitive results, better numerical schemes are required to further investigate the problem for smaller values of $\epsilon$ and larger times. 
This is currently under our investigation. 

\section*{Acknowledgement}
Research of GL \& IV was partially supported by the National Science Foundation under grant number DMS-0845127. The authors would like to thank Ken McLaughlin, Peter Miller, and Jared Bronski for useful conversations about the semiclassical limit of the focusing NLS equation. The authors would also like to thank the anonymous referee for a very thorough and helpful review. 

\appendix
\section{Background: Inverse-Scattering Transform}\label{sec:ist}
In this appendix, for the benefit of readers who may not be familiar with the inverse-scattering transform, we outline its relevant aspects here. Notably, we also outline the origin of the linear system that we solve to generate members of the Gaussian SSE; further details on these calculations can be found in \cite{MK,LM}. The procedure for solving the nonlinear initial-value problem  \eqref{eq:nls_ivp} via IST is analogous to the procedure of solving initial-value problems for linear partial differential equations by Fourier transform. Briefly, the initial data is mapped to the scattering data; the scattering data have a simple evolution in time; and the solution at later times is reconstructed from the time-evolved scattering data.  This discussion is modeled after \cite{M_notes}. For more details, see, e.g., \cite{APT,FT,ZS}.

The Lax pair for the focusing NLS equation \eqref{eq:nls} consists of the following two linear equations 
\beq
\eps\frac{\pd \vec{w}}{\pd x}=\begin{bmatrix} -\mi\lam & u \\ -u^* & \mi\lam\end{bmatrix}\vec{w}=:\mat{U}\vec{w}
\label{eq:zsa}
\eeq
and 
\beq\label{eq:lp2}
\eps\frac{\pd \vec{w}}{\pd t}=\begin{bmatrix} -\mi\lam^2+\frac{\mi}{2}|u|^2 & \lambda u+\frac{\mi}{2}\pd_xu \\ -\lam u^*+\frac{\mi}{2}\pd_x u^* & \mi\lam^2-\frac{\mi}{2}|u|^2\end{bmatrix}\vec{w}=:\mat{V}\vec{w}\,.
\eeq
Evidently, the zero-curvature condition
\beq\label{eq:zc}
\frac{\pd\mat{U}}{\pd t}-\frac{\pd\mat{V}}{\pd x}+[\mat{U},\mat{V}]=\mat{0}
\eeq 
is equivalent to the NLS equation. Note that the left-hand side of \eqref{eq:zc} is independent of \lam\ and vanishes exactly when $u$ solves \eqref{eq:nls}. It is precisely the existence of the Lax pair that allows the construction of a large family of exact solutions. Effectively, we are able to replace the nonlinear problem \eqref{eq:nls} with the pair of linear problems \eqref{eq:zsa}, \eqref{eq:lp2}. 

\subsection{Scattering}
The first step is a careful study of the problem \eqref{eq:zsa} for $\lam\in\RR$ with $u=u_0(x)$. The analysis is facilitated by the fact that $|u_0(x)|$ decays rapidly as $|x|\to\infty$, whence $\mat{U}$ tends to the constant matrix $-\mi\lam\sigma_3$ in the limit. It is precisely this observation that allows one to construct particular solutions, the Jost solutions, of the linear system \eqref{eq:zsa} normalized at $\pm\infty$. The Jost solution matrices $\mat{J}_\spm(x;\lam)$ (the  $2\times 2$ matrices whose columns are the normalized Jost solutions) are both nonsingular fundamental matrices for the differential equation. That is, 
\[
\frac{\pd\mat{J_\spm}}{\pd x}=\mat{U}\mat{J_\spm}\,.
\]
Of course, the $2\times 2$ system can only have two linearly independent column solutions, and therefore there is a $2\times 2$ matrix $\mat{S}$, called the scattering matrix, such that $J_\sp(x;\lam)=\mat{S}(\lam)\mat{J}_\sm(x;\lam)$. Further analysis reveals that the scattering matrix can be written in the form
\beq
\mat{S}(\lam)=\begin{bmatrix} a(\lam)^* & b(\lam)^* \\ -b(\lam) & a(\lam) \end{bmatrix}\,,\quad\lam\in\RR\,.
\eeq
Here, $a$ and $b$ are complex-valued functions, and they form the basis of the transmission coefficient $T(\lam)=1/a(\lam)$ and the reflection coefficient $R(\lam)=b(\lam)/a(\lam)$.

Now, it turns out that the function $a$ has an analytic continuation to the upper-half of the complex plane. Indeed, a careful look at the definition of $a$---it is a determinant whose columns are formed from Jost solutions one decaying at each of the spatial infinities---shows that zeros of $a$ in the upper half plane are $L^2(\RR)$ eigenvalues of \eqref{eq:zsa}. Associated with each such eigenvalue $\lambda_k$, there is a complex number $\gamma_k$ which is the ratio of the two analytic solutions of \eqref{eq:zsa} which make up the Wronskian $a$. The reflection coefficient $R$ does not generally extend off of the real line.

\subsection{Time evolution}

Now, if $u(x,t)$ solves \eqref{eq:nls} with initial data $u_0(x)$, then for each $t>0$ the entries in the coefficient matrix $\mat{U}$ will change. This means that the eigenvalues $\{\lambda_k\}$, the associated proportionality constants $\{\gamma_k\}$ and the reflection coefficient could be computed independently for each positive time. However, equation \eqref{eq:lp2} constrains the temporal evolution, and it is possible to write down explicit formulae which describe the time evolution. In particular, the Jost matrices, which we now write as $\mat{J}_\spm(x,t;\lam)$, must satisfy the differential equation
\[
\frac{\pd\mat{J}_\spm}{\pd t}=\mi\lam^2\mat{J}_\spm\sigma_3+\mat{V}\mat{J}_\spm\,.
\]
This, in turn, is enough to derive a differential equation in $t$ for the scattering matrix $\mat{S}(\lam;t)$:
\beq\label{eq:sode}
\frac{\pd\mat{S}}{\pd t}=\mi\lam^2[\mat{S},\sigma_3]\,.
\eeq 
Writing \eqref{eq:sode} in components, we immediately discover that when $u(x,t)$ satisfies \eqref{eq:nls}, the function $a(\lam;t)=a(\lam)$ is independent of $t$, and 
\[
b(\lam;t)=b(\lam;0)\me^{2\mi\lam^2 t}\,.
\]
It is immediate that the eigenvalues $\{\lam_k\}$ are independent of $t$ and that the reflection evolves simply as $R(\lam;t)=R(\lam;0)\me^{2\mi\lam^2t}$. Finally, a brief calculation shows that $\gamma_k(t)=\gamma_k(0)\me^{2\mi\lam^2t}$.

\subsection{Inverse scattering}
The solution $u(x,t)$ can be recovered from the scattering data (eigenvalues, proportionality constants, reflection). One way to visualize this process is to combine the columns of the Jost matrices to obtain matrices which extend into the half planes $\Im\lam\gtrless 0$. These matrices are meromorphic functions of $\lambda$ on the disjoint half planes with (generically) simple poles at the $\lam_n$'s and their complex conjugates. The residues at these poles can be computed.  Moreover, the boundary values of these matrices do not generally match on the real line; their mismatch can be quantified in terms of the reflection coefficient. Finally, the large-$\lam$ asymptotics of these matrices are prescribed. Moreover, the solution $u$ of the NLS equation is encoded in the large-$\lam$ behavior of the second column of this matrix. The inverse scattering process amounts to turning this process on its head; the properties of the matrix enumerated above (meromorphicity on $\Im\lam\gtrless 0$, prescribed poles and residues, prescribed jump across $\RR$, prescribed large-$\lam$ behavior) are in most cases sufficient to determine the matrix itself. A convenient way to organize this information is in a Riemann--Hilbert problem as follows. 

\begin{rhp}\label{rhp:2}
Find a $2\times 2$ matrix-valued function $\mat{m}(\lambda;x,t)$ with the following properties.
\begin{enumerate}
\item $\mat{m}(\lambda;x,t)$ is an analytic function on 
\[
\CC\setminus(\RR\cup\{\lambda_0,\lambda_1,\ldots,\lambda_{N-1},\lambda_0^*,\ldots,\lambda_{N_1}^*\})\,.
\]
\item $\mat{m}(\lambda;x,t)\to\Id$ as $\lambda\to\infty$.
\item  $\mat{m}$ has simple poles at the points $\lambda_{k}$ and $\lambda_{k}^*$; the residues satisfy: 
\begin{align}
\res_{\lambda=\lambda_{k}}\mat{m}(\lambda)& =\lim_{\lambda\to\lambda_{k}}\mat{m}(\lambda)\begin{bmatrix}   0 & 0 \\ e_{k}(x,t) & 0\end{bmatrix}\,, \\
\res_{\lambda=\lambda_{k}^*}\mat{m}(\lambda)& =\lim_{\lambda\to\lambda_{k}^*}\mat{m}(\lambda)\begin{bmatrix}   0 & -e_{k}(x,t)^* \\ 0 & 0\end{bmatrix}\,.
\end{align}
Here, $e_{k}(x,t)$ is given explicitly in terms of the $\gamma_k$'s.
\item The matrix $\mat{m}(\lam;x,t)$ takes continuous boundary values on $\RR$. For $\lam\in\RR$ we write 
\[
\mat{m}_\spm(\lam;x,t):=\lim_{\delta\downarrow 0}\mat{m}(\lam\pm\mi\delta;x,t)\,,
\]
and 
\beq
\mat{m}_\sp(\lam;x,t)=\mat{m}_\sp(\lam;x,t) \mat{v}(\lambda;x,t)
\eeq
with the jump matrix $\mat{v}$ given explicitly in terms of the reflection $R$.
\end{enumerate}
Finally, once the solution of RHP \ref{rhp:2} is found, one recovers the solution via the formula
\[
u(x,t)=2\mi\lim_{\lambda\to\infty}\lambda m_{12}(\lambda;x,t)\,.
\]
\end{rhp}

\subsection{$N$-soliton solutions, linear system}

A solution for which $R(\lam)\equiv0$ is completely characterized by the $N$ eigenvalues in the upper half plane (and their proportionality coefficients). In this case the jump matrix satisfies $\mat{v}=\Id$ so there is no mismatch between $\mat{m}_\spm$ on $\RR$. This is the case we consider in this paper. 

If we make the partial-fractions ansatz
\beq
\mat{m}(\lambda;x,t)=\Id+\sum_{k=0}^{N-1}\frac{\mat{A}_k(x,t)}{\lambda-\lambda_{k}}+\sum_{k=0}^{N-1}\frac{\mat{B}_k(x,t)}{\lambda-\lambda_{k}^*}\,,
\eeq
it's clear from RHP \ref{rhp:1} that
\beq
\res_{\lambda=\lambda_{k}}\mat{m}(\lambda;x,t)=\mat{A}_k(x,t)
\eeq
and 
\beq
\res_{\lambda=\lambda_{k}^*}\mat{m}(\lambda;x,t)=\mat{B}_k(x,t)\,.
\eeq
Moreover, we see that the matrices $\mat{A}_k$ must have zeros in the second column,
\beq
\mat{A}_k(x,t)=\begin{bmatrix} a_k(x,t) & 0 \\ b_k(x,t) & 0 \end{bmatrix}\,,
\eeq 
while the matrices $\mat{B}_k$ have zeros in the first column
\beq
\mat{B}_k(x,t)=\begin{bmatrix} 0 & c_k(x,t) \\ 0 & d_k(x,t) \end{bmatrix}\,,
\eeq 
Then
\beq\label{eq:linear1}
\begin{bmatrix} a_k \\ b_k \end{bmatrix} = e_{k}\left(\begin{bmatrix} 0 \\ 1 \end{bmatrix} +\sum_{j=0}^{N-1}\frac{1}{(\lambda_{k}-\lambda_{j}^*)}\begin{bmatrix} c_j \\ d_j \end{bmatrix}\right)\,.
\eeq
Similarly, 
\beq\label{eq:linear2}
\begin{bmatrix} c_k \\ d_k \end{bmatrix} = e_{k}^*\left(\begin{bmatrix} 1 \\ 0 \end{bmatrix} +\sum_{j=0}^{N-1}\frac{1}{(\lambda_{k}^*-\lambda_{j})}\begin{bmatrix} a_j \\ b_j \end{bmatrix}\right)\,.
\eeq
Evidently, \eqref{eq:linear1}, \eqref{eq:linear2} form a linear system for the coefficients in $\mat{A}_k$ and $\mat{B}_k$.

\bibliographystyle{plain}
\bibliography{llv}
\end{document}